\documentclass{article}
\usepackage{graphicx} 
\usepackage{authblk}
\usepackage[margin=1in]{geometry}
\usepackage{amsmath, amssymb}
\usepackage{dirtytalk}
\usepackage{hyperref}
\usepackage[english]{babel}
\usepackage[toc,page]{appendix}
\usepackage{subcaption}
\usepackage[x11names, svgnames, dvipsnames]{xcolor}
\usepackage{wrapfig}
\usepackage{dirtytalk}
\usepackage{lineno}
\usepackage[T1]{fontenc}
\usepackage{tikz}
\usepackage{cite}
\usepackage{float}
\usepackage{enumitem}
\usetikzlibrary{trees}

\newtheorem{definition}{Definition}

\title{Towards Quantum Tensor Decomposition in  Biomedical Applications}

\author[1]{Myson Burch}
\author[2,3]{Jiasen Zhang}
\author[4]{Gideon Idumah}
\author[5]{Hakan Doga}
\author[3,6]{Richard Lartey}
\author[7]{Lamis Yehia}
\author[3,6]{Mingrui Yang}
\author[8]{Murat Yildirim}
\author[7,10]{Mihriban Karaayvaz}
\author[9]{Omar Shehab}
\author[2]{Weihong Guo}
\author[4,10,11]{Ying Ni}
\author[1]{Laxmi Parida}
\author[3,6,12]{Xiaojuan Li}
\author[1]{Aritra Bose\thanks{Correspondence: a.bose@ibm.com}}

\affil[1]{IBM Research, Yorktown Heights, NY}
\affil[2]{Department of Mathematics, Case Western Reserve University, Cleveland, OH}
\affil[3]{Program for Advanced Musculoskeletal Imaging (PAMI), Cleveland Clinic, Cleveland, OH}
\affil[4]{Center for Immunotherapy and Precision Immuno-Oncology, Lerner Research Institute, Cleveland Clinic, Cleveland, OH}
\affil[5]{IBM Quantum, Almaden Research Center, San Jose, CA}
\affil[6]{Department of Biomedical Engineering, Lerner Research Institute, Cleveland Clinic, Cleveland, OH}
\affil[7]{Genomic Medicine Institute, Lerner Research Institute, Cleveland Clinic, Cleveland, OH}
\affil[8]{Department of Neurosciences, Lerner Research Institute, Cleveland Clinic, Cleveland, OH}
\affil[9]{IBM Quantum, Yorktown Heights, NY, USA}
\affil[10]{Department of Molecular Medicine, Cleveland Clinic Lerner College of Medicine, Case Western Reserve University, Cleveland, OH}
\affil[11]{Case Comprehensive Cancer Center, Case Western Reserve University, Cleveland, OH}
\affil[12]{Department of Biomedical Engineering, Case Western Reserve University, Cleveland, OH}

\date{}

\begin{document}

\maketitle

\begin{abstract}
   Tensor decomposition has emerged as a powerful framework for feature extraction in multi-modal biomedical data. In this review, we present a comprehensive analysis of tensor decomposition methods such as Tucker, CANDECOMP/PARAFAC, spiked tensor decomposition, etc. and their diverse applications across biomedical domains such as imaging, multi-omics, and spatial transcriptomics. To systematically investigate the literature, we applied a topic modeling-based approach that identifies and groups distinct thematic sub-areas in biomedicine where tensor decomposition has been used, thereby revealing key trends and research directions. We evaluated challenges related to the scalability of latent spaces along with obtaining the optimal rank of the tensor, which often hinder the extraction of meaningful features from increasingly large and complex datasets. Additionally, we discuss recent advances in quantum algorithms for tensor decomposition, exploring how quantum computing can be leveraged to address these challenges. Our study includes a preliminary resource estimation analysis for quantum computing platforms and examines the feasibility of implementing quantum-enhanced tensor decomposition methods on near-term quantum devices. Collectively, this review not only synthesizes current applications and challenges of tensor decomposition in biomedical analyses but also outlines promising quantum computing strategies to enhance its impact on deriving actionable insights from complex biomedical data. 
\end{abstract}


\section{Introduction} 

Biomedical data encompasses a diverse set of multi-modal and multi-omics data from healthcare and life sciences providing critical insights into the complexity of disease etiology and biological functions. A comprehensive understanding of biomedical data is required to advance precision medicine and treatment strategies for complex diseases. Innovations in technologies like genomics, proteomics, medical imaging, such as magnetic resonance imaging (MRI), computed tomography (CT), etc. have resulted in high-quality data that enable personalized clinical decision making. This opportunity also presents us with significant challenges in data management and analysis~\cite{lambin2017radiomics}, such as representing the different modalities of biomedical data and finding higher-order relationships between them. With the increasing size and complexity of biomedical data, tensors  are emerging as a powerful method for the integration and analysis of data ~\cite{ho2014marble,luo2017tensor,wang2021variational, jung2021monti, taguchi2022novel, gao2023biostd, liu2023multiomics, zhou2013tensor}.

Tensors are multidimensional arrays rendered as generalizations of vector spaces~\cite{kolda2009tensor}. A vector is a first-order tensor, a matrix is a second-order tensor, and tensors of order three or higher are called higher-order tensors. Tensor decompositions, although originally proposed about a century ago~\cite{hitchcock1927expression}, have found renewed interest in several applications in signal processing, machine learning~\cite{sidiropoulos2017tensor}, data mining~\cite{papalexakis2016tensors}, computer vision~\cite{shashua2005non}, neuroscience~\cite{beckmann2005tensorial}, etc. in the past two decades, with the ability to represent complex and multimodal data. Tensors have the ability to represent each biomedical modality in a different dimension. Hence, they can be used to integrate multi-modal biomedical data spanning multi-omics variables. Tensor factorization or decomposition allows us to obtain a latent structure from a higher-dimensional space spanned by all the multi-modal variables and identify interactions between them~\cite{kolda2009tensor, luo2017tensor}. Tensor decomposition (TD) has widespread application in biomedical data analysis ranging from analyzing MRI images and electroencephalography (EEG) signals~\cite{sedighin2024tensor}, finding genotype-phenotype relationships~\cite{kessler2014learning}, quantifying expression quantitative trait loci (eQTL)~\cite{hore2016tensor}, integrating multi-omics data~\cite{jung2021monti}, to spatial transcriptomics~\cite{broadbent2024deciphering}. There have been attempts to review the applications of TD in biomedical data but they were focused on specific applications such as imaging analysis~\cite{sedighin2024tensor}, precision medicine~\cite{luo2017tensor} or, on specific diseases such as cancer~\cite{movahed2024tensor}, and often limited in scope. 

This review examines recent advances in tensor decomposition for biomedical data analysis. We review its application across diverse biomedical domains over the past decade, highlighting state-of-the-art practices, the inherent advantages of these methods, and the scalability challenges that remain. Moreover, we discuss a unique perspective of adapting emerging technologies such as quantum computing (QC) and strategies for developing quantum algorithms for TD to ameliorate the challenges arising in classical TD applications. QC has the potential to accelerate computing in biomedicine~\cite{durant2024primer, flother2024quantum, emani2021quantum, nalkecz2024quantum}, with possible applications in single-cell analysis~\cite{basu2023towards,utro2024perspective}, predictive modeling in multi-omics analysis~\cite{bose2024quantum}, clinical trials~\cite{doga2024can}, breast cancer subtyping using histopathological images~\cite{ray2024hybrid}, protein structure prediction~\cite{doga2024perspective, raubenolt2023quantum}, among others. We explore development of a quantum tensor decomposition (QTD) algorithm and provide a framework of how it can be implemented in present-day pre-fault tolerant quantum devices (PFTQD) to solve problems in biomedicine, as well as discussing future implementations in fault-tolerant quantum devices.

This review is structured as follows. We introduce different types of TD approaches in Section~\ref{sec:td_def} and discuss methods to evaluate computational hardness in Section~\ref{sec:hardness}. Next, we continue to review the current literature for TD applications in biomedicine using an unsupervised technique in Section~\ref{sec:td_biomed} and subsequently discuss the main findings on imaging (Section~\ref{sec:td_imaging}), multi-omics (Section~\ref{sec:td_multiomics}), and spatial transcriptomics (Section~\ref{sec:td_spatial}). We also analyze the hardness of tensor decomposition in real-world MRI and multi-omics data, highlighting the factors impacting its implementation in classical systems in Section~\ref{sec:challenges}. Thereafter, we review the literature on quantum algorithms for TD (Section~\ref{sec:td_quantum}) and discuss particular use cases where quantum tensor decomposition can be applied in biomedical data (Section~\ref{sec:td_usecase}). Finally, we summarize our review and future perspective in Section~\ref{sec:conclusion}.

\section{A Primer on Tensor Decomposition}~\label{sec:td_def}
Mathematically, tensors are defined as multilinear functions on the Cartesian product of vector spaces \cite{spivak2018calculus}; however, it is helpful to think of tensors as a generalization of matrices to higher dimensions. As a result, tensor decompositions can be viewed as a generalization of matrix decomposition to higher-orders. The \textit{order} of a tensor is usually defined as the number of its dimensions, or equivalently the number of vectors spaces included in the product. As decomposing matrices to lower rank is generally non-unique~\cite{rabanser2017introduction} and only unique under certain constraints, such as imposed by the singular value decomposition (SVD)~\cite{kolda2009tensor}, tensor decompositions in higher-order are shown to be unique with minimal constraints and hence more general.

We will start by defining notation and preliminary properties of tensors (for a more detailed introduction to tensor decomposition methods, we refer to~\cite{kolda2009tensor}). Scalars are denoted in lowercase letters, e.g., $a$. We denote vectors (tensors of order one) in bold lowercase such as $\mathbf{u}$, with its $i^{th}$ element being $u_i$. Matrices (order-two tensors) are denoted in bold uppercase, e.g. $\mathbf{X}$ and element $(i,j)$ is denoted as $x_{ij}$. The $n^{th}$ element in a sequence is denoted as $\textbf{u}^{(n)}$, such that it is the $n^{th}$ vector in a sequence of vectors. Higher-order tensors (order three or more) are defined in Euler script, e.g., $\mathcal{T}$. 

Next, we define important properties of tensors such as its \textit{mode}, which is another name for the \textit{order} of the tensor (also called \textit{way} in some cases).  It is a measure of how many indices exist in the tensor. A zero tensor of any order is defined as a tensor with all its elements equal to zero. A third-order (or 3-way) tensor $\mathcal{T}$ has three indices $T \in \mathbb{R}^{I\times J\times K}$, where $I,J,K$ are the three indices respectively. 
Another important property of tensors are \textit{slices}, which are two-dimensional sections of a tensor, defined by fixing all but two indices. For a third-order tensor $\mathcal{T}$, we will have horizontal, lateral, and frontal slices denoted by $\mathcal{T}_{i::}$, $\mathcal{T}_{:j:}$, and $\mathcal{T}_{::k}$, respectively~\cite{kolda2009tensor}. Moreover, we use the notation ``$\circ$'' to the denote outer product of vectors, and adopt ``$\otimes$'' for the tensor product operation. While these two operations are different and yield different mathematical objects, they are related in the sense that the result of a tensor product can be viewed as the vectorization of an outer product output. Next, we define rank-one tensors and the rank of a tensor before defining different types of tensor decompositions. 


\begin{definition}~\label{def:rankone}
   \textbf{(Rank-one tensor).} A $d$-way non-zero tensor $\mathcal{T} \in \mathbb{R}^{N_1 \times N_2 \times \cdots \times N_d}$ is of rank $one$ if and only if it can be written as 
    $$
        \mathcal{T} = \mathbf{u}^{(1)} \circ \mathbf{u}^{(2)} \circ \cdots \circ \mathbf{u}^{(d)}, 
    $$
    where $\mathbf{u}^{(i)} \in \mathbb{R}^{N_i}$ and $\circ$ denotes the outer product. (In other words, each entry $t_{n_1,n_2,\cdots,n_d}$ of the tensor is the product, $u_{n_1}^{(1)}u_{n_2}^{(2)}\cdots u_{n_d}^{(d)}$, of the corresponding vector entries).
\end{definition}

\begin{definition}~\label{def:rank}
    \textbf{(Tensor rank).} The $rank$ of a tensor $\mathcal{T}$, denoted as \texttt{rank}($\mathcal{T}$) is defined as the smallest integer $r$ such that there exist $r$ rank one tensors $\mathcal{T}_1, ..., \mathcal{T}_r$ with
    $$
        \mathcal{T} = \sum_{i=1}^{r} 
        \mathcal{T}_i.
    $$
\end{definition}

The idea of decomposing a tensor into its polyadic form, i.e. expressing it as the sum of a finite number of rank-one tensor has existed for over a century~\cite{hitchcock1927expression}. It has been re-introduced in the literature as CANDECOMP (canonical decomposition) and PARAFAC (parallel factors) later and collectively known as the CP decomposition~\cite{harshman1970foundations, harshman1972parafac2}. It factorizes a tensor into a sum of component rank-one tensor as defined in Definition~\ref{def:rankone}. 
\begin{figure}[!htbp]
    \centering
    \includegraphics[width=0.8\linewidth]{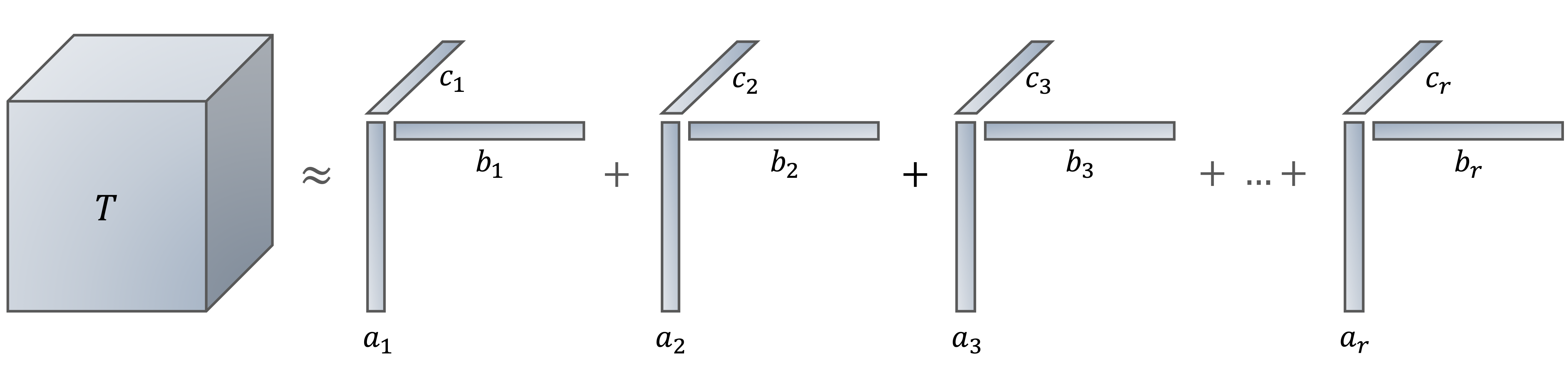}
    \caption{CP decomposition of a third-order tensor into $r$ rank-one tensors}
    \label{fig:cp}
\end{figure}
    
    As an example, a third-order CP decomposition (Figure~\ref{fig:cp}) can be written as 
    \begin{equation}\label{eq:cp}
        \mathcal{T} \approx \sum_{i=1}^r \lambda_i \, \mathbf{u}_i^{(1)} \circ \mathbf{u}_i^{(2)} \circ \cdots \circ \mathbf{u}_i^{(d)},
    \end{equation}
    where $\mathbf{u}_i^{(n)} \in \mathbb{R}^{N_d}$ are the factor vectors for order $d$, $\lambda_i$ is the scaling weight, and  $r$ is the tensor rank. 

\subsection{Hierarchy of tensor decompositions}
\begin{wrapfigure}{l}{0.5\textwidth}
      \centering
        \includegraphics[width=0.48\textwidth]{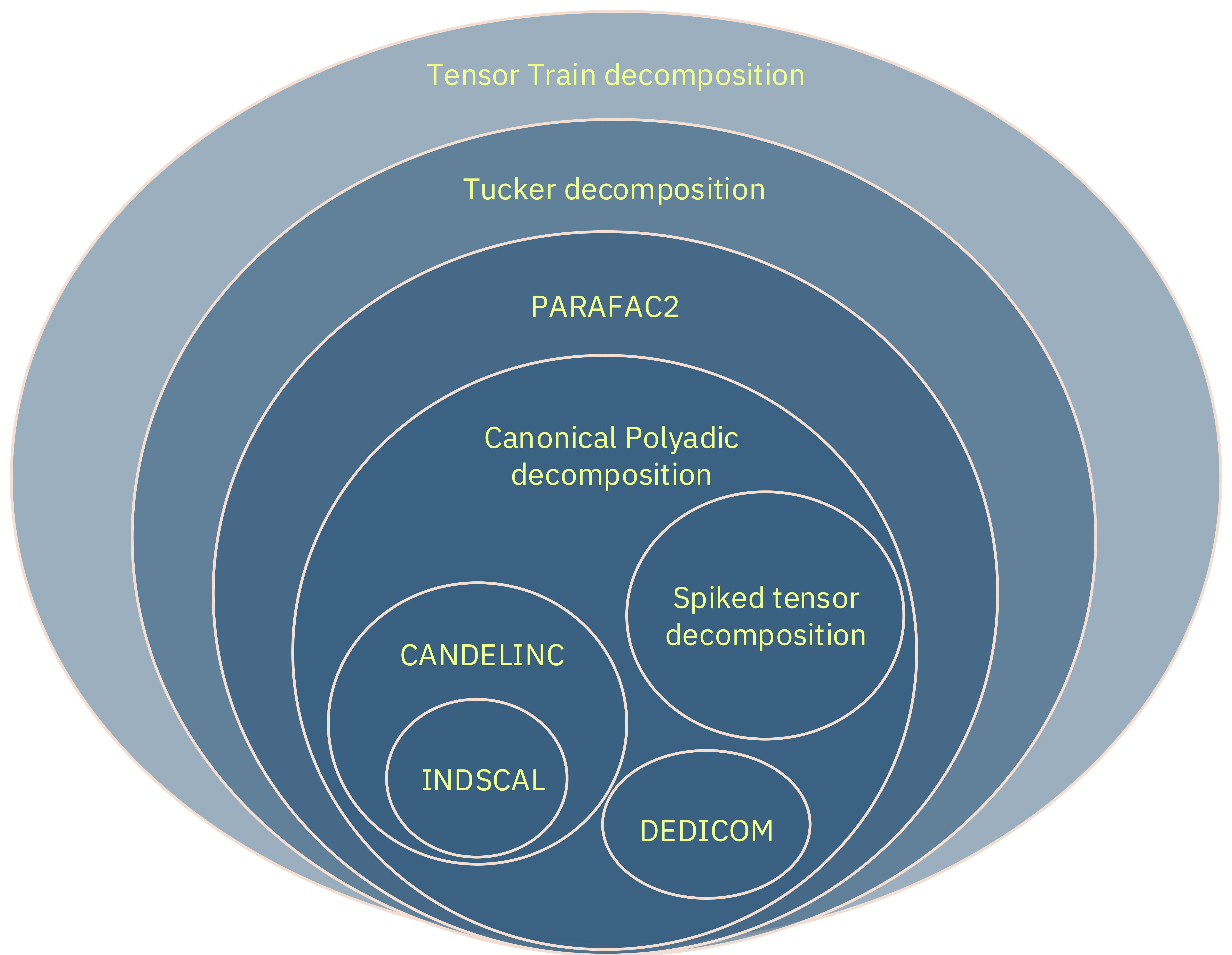}
    \caption{Hierarchy of tensor decompositions. (See Sections ~\ref{sec:indscal-to-candelinc} for mapping of INDSCAL onto CANDELINC, ~\ref{sec:dedicom-to-parafac} for mapping of DEDICOM onto PARAFAC.)}
    \label{fig:hierarchy}
\end{wrapfigure}

The hierarchy of TD methods \cite{kolda2009tensor} reveals a nested structure in which broader frameworks encompass more specialized approaches (Figure~\ref{fig:hierarchy}). Therefore, any computational advantage for a specific tensor decomposition applies to all of its subtypes. At the top of this hierarchy is Tensor Train (TT) decomposition. It represents a higher-order tensor as a sequential \say{train} of lower-order tensors (TT cores), each connected through shared dimensions known as TT ranks~\cite{oseledets2011tensor}. 
\begin{definition}\label{def:tt}
    \textbf{(Tensor Train decomposition).}  Given a $d$-order tensor $\mathcal{T} \in \mathbb{R}^{N_1 \times N_2 \times \dots \times N_d}$, TT decomposition factorizes it into a sequence of \textit{3-way} core tensors $G^{(n)}$ (except for the first and last cores, which are matrices), connected in a train-like structure as a product of $d$ core entries 
    as follows,
    \begin{equation}\label{eq:tt}
        \mathcal{T}(n_1, n_2, \cdots, n_d) = \sum_{r_0, r_1, \dots, r_{d-1},r_d}G^{(1)}(r_0,n_1,r_1) G^{(2)}(r_1,n_2,r_2) \cdots  G^{(d)}(r_{d-1},n_d,r_d) 
    \end{equation}
    where $G^{(n)} \in \mathbb{R}^{R_{n-1}\times N_n \times R_n}$ for $1 \leq n \leq d$. $G^{(1)} \in \mathbb{R}^{N_1 \times R_1}$ is a matrix in the first mode and $G^{(d)} \in \mathbb{R}^{R_{d-1} \times N_d}$ is a matrix for the last mode. Here $R_n$ are \texttt{TT-ranks} that control the compression and approximation quality. 
\end{definition}


The computational complexity for constructing the TT decomposition using methods like TT-SVD is $\mathcal{O}(dN^d)$ for unstructured tensors. However, it becomes manageable in practice for low TT ranks and structured tensors, making it highly effective for large-scale, high-dimensional data applications \cite{oseledets2011tensor, Zhang2020trillion}.

A special case of TT decomposition is Tucker Decomposition~\cite{tucker1966some, zniyed2020high}. It is a tensor factorization method that expresses a high-order tensor as a core tensor multiplied by factor matrices along each mode. 

\begin{definition}\label{def:tucker}
    \textbf{(Tucker decomposition).} Given a $d$-order tensor $\mathcal{T} \in \mathbb{R}^{N_1 \times N_2 \times \cdots \times N_d}$, the decomposition is represented as, 
    \begin{equation}\label{eq:tucker}
        \mathcal{T} = \mathcal{G} \times_1 U^{(1)} \times_2 U^{(2)} \cdots \times_d U^{(d)},
    \end{equation}
     $\mathcal{G} \in \mathbb{R}^{T_1 \times T_2 \times \cdots \times T_d}$ is the core tensor, and $U^{(n)} \in \mathbb{R}^{N_n \times T_n}$ are factor matrices and $\times_n$ denotes $n$-mode matrix multiplication. The multilinear ranks $T_n$ determine the dimensionality of the core tensor along each mode.
\end{definition}

Tucker decomposition is often computed using algorithms like High-Order Singular Value Decomposition (HOSVD). The storage cost is $\mathcal{O}(dNT + T^d)$, where $N = \max(N_n)$, $T = \max(T_n)$, and $T^d$ reflects the size of the core tensor. The computational complexity of HOSVD is $\mathcal{O}(dN^{d-1}T)$, dominated by the SVD of the mode unfoldings of the tensor. Tucker decomposition is widely used for dimensionality reduction, compression, and exploratory analysis of multi-way data~\cite{kolda2009tensor}.

Unlike Tucker decomposition, CP imposes a more constrained structure, making it unique under mild conditions and suitable for identifying latent components. The storage complexity is $\mathcal{O}(rdN)$, where $N = \max(N_d)$, the order $d \ll N$, and rank $r$. Computational complexity depends on iterative algorithms like Alternating Least Squares (ALS), requiring $\mathcal{O}(rN^d)$ operations per iteration for unstructured tensors, but can be reduced with sparse or structured data. CP is extensively used in fields like chemometrics, psychometrics, and machine learning for data interpretation and factorization~\cite{kolda2009tensor}. 


PARAFAC2 is a generalization of CP decomposition that relaxes the constraint of fixed factors and is applicable to a collection of matrices where each have the same number of columns but, varying number of rows. PARAFAC2 applies the same factor across one mode while allowing the other factor matrix to vary, rendering it suitable for real-world scenarios where one mode varies in size or structure across slices \cite{harshman1972parafac2}. Like CP, it expresses a tensor as a sum of rank-one components. However, PARAFAC2 introduces slice-specific factor matrices \( \mathbf{B}_k \) for the varying mode, with the constraint \( \mathbf{B}_k^\top \mathbf{B}_k = \mathbf{B}_j^\top \mathbf{B}_j \) for all \( k, j \). 
\begin{definition}\label{def:parafac2}
    \textbf{PARAFAC2 decomposition.}  The decomposition is given by
    $$
    \mathcal{X}_k = \mathbf{A} \, \text{diag}(\mathbf{c}_k) \, \mathbf{B}_k^\top,
    $$
    where $\mathbf{A}$ is shared across slices, $\mathbf{c}_k$ contains slice-specific weights, and $\mathbf{B}_k$ adapts to the varying dimensions.  
\end{definition}

 The storage complexity of PARAFAC2 is $\mathcal{O}(rN + r\sum_k J_k )$, where $N$ is the size of the first mode, $J_k$ the size of the varying mode, and $r$ the rank. Computational complexity depends on iterative methods like ALS, typically requiring $\mathcal{O}(rN \max(J_k) K)$ operations per iteration for $K$ slices. PARAFAC2 is widely used for longitudinal or time-varying data in applications like recommender systems and signal processing. PARAFAC2 decomposition is a specialization of Tucker decomposition (See Appendix for a detailed example). 

 Within PARAFAC2, the CP decomposition emerges as a more constrained subset, representing tensors as a sum of rank-one components (Definition~\ref{def:rankone}). Furthermore, spiked tensor decomposition, CANDELINC, and DEDICOM tensor decompositions are subsets of CP decomposition, designed for specific applications or structural assumptions. Finally, INDSCAL tensor decomposition is a refined version of CANDELINC, incorporating individual differences scaling to model individual variation in tensor components. 
 
 The spiked tensor decomposition requires structural assumptions on the data such as the existence of a low-rank signal tensor in presence of noise. The assumption is that of a \say{spiked} structure, which separates the underlying signal from the noise. It is a specific case of CP decomposition that, like CP, recovers a low-rank structure of the tensor. However, when the signal-to-noise ratio (SNR) is low, it resembles the CP decomposition. 
 \begin{definition}\label{def:st}
     \textbf{Spiked Tensor decomposition.} An underlying statistical model called the \say{spiked tensor model}, introduced in \cite{richard2014statistical}, is considered, where a $N$-dimensional real or complex signal vector $v_{sig}$ is randomly chosen to form 
     \begin{equation}\label{eq:st}
          T_0 = \lambda v_{sig}^{\otimes p} +G
     \end{equation} 
       
    In this spiked tensor formulation, $\lambda$ represents the signal strength or SNR and $G$ is Gaussian noise. The task is either to approximate the signal vector $v_{sig}$ or to determine whether the signal can be detected above some predetermined threshold.
 \end{definition}

The hierarchical framework of tensor decompositions, based on their computational complexity classes (Figure~\ref{fig:hierarchy}), highlights the interrelatedness of TD methods while showcasing their suitability for distinct scenarios and data characteristics.

\subsection{Phase transition of computational hardness}~\label{sec:hardness}
A computational problem is said to exhibit a \textit{phase transition} if there exists a sharp threshold in some control parameter, such as signal-to-noise ratio or rank, at which the problem changes from being computationally feasible (solvable in polynomial time) to infeasible (requiring super-polynomial or exponential time). Mathematically, let $P(n)$ be a problem instance of size $n$ with a control parameter $\theta(n)$. The computational phase transition occurs at some critical threshold $\theta_c(n)$, such that:

\[
\lim_{n \to \infty} \mathbb{P}(\text{efficient algorithm solves } P(n)) = \begin{cases} 1, & \theta(n) < \theta_c(n) \\ 0, & \theta(n) > \theta_c(n) \end{cases}
\]

where \say{efficient algorithm} refers to an algorithm with polynomial runtime in $n$.

Various phase transition parameters are available to study the computational hardness in tensor decomposition. A commonly used hardness metric to quantify computational tractability is \textit{Minimum Mean Squared Error} (MMSE) in low-degree polynomials~\cite{wein2023average}. In a random low-rank tensor decomposition, if the tensor rank satisfies $r \ll n^{k/2}$, then $\text{MMSE} \to 0$ and the decomposition is considered computationally feasible with efficient signal recovery . Otherwise, if it is $r \gg n^{k/2}$, then $\text{MMSE} \to 1$ and the decomposition is infeasible without efficient recovery~\cite{wein2023average}. This threshold applies to general order-$k$ tensors and provides a tight characterization of computational hardness.


Signal strength is also used as a parameter for phase transition. In higher-order tensor clustering, the decay rate of the signal, $\beta$ controls the phase transition in the form of $\tilde{\Theta}(n^{-\beta})$~\cite{luo2022tensor}. Specific values of $\beta$ indicates computationally easy and hard regimes. 
Similarly, for sparse tensor decomposition, sparsity is a crucial parameter. It is defined as 

\[
k = \tilde{\Theta}(n^{\alpha})
\]

where $\alpha$ represents the sparsity in the form of the fraction of non-zero entries in the tensor. Lower values of $\alpha$ makes the tensor decomposition computationally harder~\cite{luo2022tensor}. The Signal-to-Noise Ratio (SNR) is another commonly used parameter for measuring hardness. In a low-rank Bernoulli model


\[
Y | \Theta \sim \text{Bernoulli}(f(\Theta)), 
\]

where $\Theta$ is a low-rank continuous-valued tensor, three  distinct phases of learnability were derived~\cite{wang2020learning}: 
\begin{itemize}
    \item \textit{High SNR (Noise Helps):} Moderate noise improves estimation due to a dithering effect.
    \item \textit{Intermediate SNR (Noise Hurts):} Standard behavior where noise degrades performance.
    \item \textit{Low SNR (Impossible Phase):} Learning is infeasible below a critical threshold.
\end{itemize}
Similarly, in a tensor completion problem, the phase transition in recovering missing values is based on \textit{Inverse SNR}, which governs the feasibility of recovery~\cite{stephan2024non}. The study of the phase transition of computational hardness  allows us to systematically investigate potential use cases for a new algorithm where the state-of-the-art approaches may not be performant.



\section{Applications in Biomedical Analysis}
~\label{sec:td_biomed}
Tensor decomposition extracts latent structures from high dimensional data and preserves interpretability of the features. This is critical in analyzing biomedical data to ensure the insights drawn from lower-dimensional factors are reliable and actionable, leading to biomarker discovery, prognosis, and developing therapeutics for complex diseases. 
To further explore the role of tensor decomposition in biomedical research, we used a topic modeling-based approach by employing BERTopic~\cite{grootendorst2022bertopic} to conduct an unsupervised analysis of the literature on tensor decomposition and biological data in PubMed. By leveraging transformer-based embeddings and topic clustering, we systematically identified key themes and research trends related to TD and its applications in biomedicine. 
The terms we searched were \say{tensor decomposition} in combination with (\texttt{AND}) the following terms: \say{genomics},
\say{transcriptomics},
\say{proteomics},
\say{metabolomics},
\say{epigenomics},
\say{microbiomics},
\say{multiomics} \texttt{OR} \say{multi-omics},
\say{cancer},
\say{cardiovascular disease},
\say{diabetes},
\say{alzheimer's disease},
\say{neurological disorder},
\say{autoimmune disease},
\say{kidney disease},
\say{obesity},
\say{medical imaging}. From those 16 terms, 536 documents were extracted from PubMed in the past ten years. This approach enabled us to identify dominant research areas, emerging applications, and potential gaps in the field, providing a structured overview of how tensor decomposition is being utilized across different biomedical domains.

\begin{figure}[htbp]
    \centering
    \includegraphics[width=\linewidth]{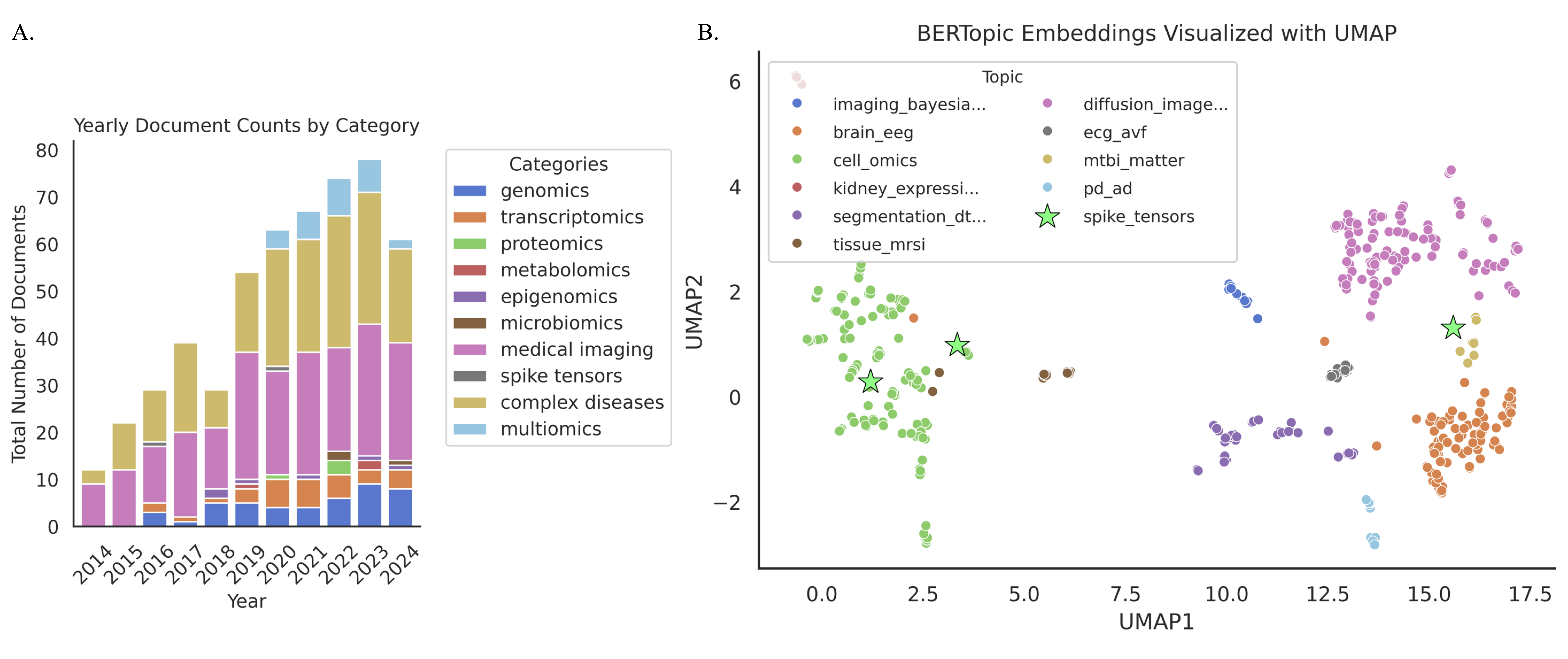}
    \caption{Tensor decomposition applications in biomedical data in the literature. \textbf{A.} Stacked bar chart of the number of papers from PubMed published with the topic \protect\say{tensor decomposition} \texttt{AND} categories listed in the legend in the past decade. \protect\textbf{B.} UMAP of the BERTopic embeddings visualizing the different topic clusters. Each data point is a research paper. Applications of \protect\say{spike tensors} are highlighted with a star symbol.}
   \label{fig:sidebyside}
\end{figure}

Our analysis revealed a well-established use of TD in medical imaging, where it has been applied for tasks such as noise reduction, feature extraction, and disease pattern recognition~\cite{bustin2019high,Brender2019,Li2021,Zhang2017,Chatzichristos2019}(Figure~\ref{fig:sidebyside}A), with 218 articles cited over 1,400 times (Supplementary Table 1). This aligns with the long history of tensor methods in signal processing and image analysis. This is followed by applications in complex diseases such as cardiovascular disease~\cite{zhao2019detecting}, with TD methods used to factor ECG signals~\cite{liu2020automated} and X-rays~\cite{sedighin2024tensor}, among other applications. Similarly, TD methods have been used to analyze EEG~\cite{cong2015tensor} and neuroimaging data for neurological diseases such as Alzheimer's disease~\cite{durusoy2018multi}, Parkinson's disease~\cite{pham2017tensor}, stroke~\cite{zhou2024tensor}, glioblastoma~\cite{yang2014discrimination}, etc. More recently, we observed a slight increase in the application of TD to multi-omics data, reflecting the growing need for integrative approaches in systems biology (Figure~\ref{fig:sidebyside}A). This trend highlights the increasing recognition of tensor methods as powerful tools for extracting meaningful patterns from high-dimensional, heterogeneous biological datasets.


To further interpret the structure of the identified topics, we visualized the document embeddings using UMAP, with topics color-coded for clarity (Figure~\ref{fig:sidebyside}B). This visualization revealed distinct topic clusters, highlighting distinct areas of research focus. The most represented topics were \say{\texttt{cell\_omics\_expression}}, \say{\texttt{diffusion\_imaging\_noise}}, and \say{\texttt{brain\_connectivity\_eeg}}, reflecting the strong presence of tensor decomposition in imaging-related applications. The \say{\texttt{cell\_omics\_expression}} topic encompassed studies leveraging tensor methods for multi-omics and spatial transcriptomics to analyze gene expression across cells and tissues~\cite{hore2016tensor,Omberg2007,Ramdhani2020,Wang2019}. The \say{\texttt{brain\_connectivity\_eeg}} topic captured research applying TD to neurological and electrophysiological data, particularly EEG and fMRI, for brain signal processing and cognitive studies~\cite{Li2021,Chatzichristos2019}. The \say{\texttt{diffusion\_imaging\_noise}} topic represented works focused on image denoising, enhancement, and reconstruction in medical imaging, further reinforcing the dominant role of tensor methods in this domain~\cite{bustin2019high,Brender2019,Li2021,Zhang2017,Chatzichristos2019}. More information on the other topic clusters are included in the Supplementary Tables 1-3.

This exploration highlights the significant role of TD in biomedicine, demonstrating its widespread application in medical imaging, multi-omics, and neuroscience. The increasing adoption of these methods emphasizes their value in extracting meaningful insights from complex biomedical data. Tensors are unique in their applications in multi-modal, multi-omics data because they have the ability to integrate these data sets, find linear and non-linear ineteractions between them, and extract meaningful latent representations which can be then used to perform downstream tasks~\cite{kolda2009tensor, luo2017tensor}. We provide a framework for integrating TD methods, including emerging methods such as quantum tensor decomposition in Figure~\ref{fig:qtd_framework}. Although insightful, this investigation is far from exhaustive and does not provide a holistic view on the literature with specific sub-areas such as genomics, MRI images, electronic health records (EHR), spatial transcriptomics, etc. having more applications of TD~\cite{hore2016tensor, olesen2023tensor, sedighin2024tensor, ho2014marble, song2023gntd}. Hence, we will focus on each of these sub-areas broadly categorized into biomedical imaging, multi-omics analysis, and spatial transcriptomics. 

\subsection{Biomedical Imaging}\label{sec:td_imaging}

Tensor decomposition methods have emerged as powerful tools for analyzing high-dimensional data in biomedical imaging~\cite{sedighin2024tensor}.
Applications span a wide range of domains, including medical image denoising, super resolution, reconstruction, and so on, where tensors effectively model spatial, temporal, and spectral relationships in MRI and hyperspectral images. Beyond MRI and conventional imaging modalities, TD techniques have shown promise in analyzing high-dimensional datasets generated by multiphoton microscopy, including fluorescence-based and label-free modalities such as second harmonic generation (SHG) and third harmonic generation (THG) imaging~\cite{jamesdarian2021recent}.  In this section, we will review these applications from a methods perspective by discussing how each tensor decomposition technique is used in biomedical imaging analysis. 

\paragraph{CP decomposition.} CP decomposition is widely used in real-world data due to its simplicity, interpretability, and ability to efficiently represent multi-dimensional data. It decomposes the data tensor into a sum of rank-one components (Equation~\ref{eq:cp}) to uncover latent structures, which is particularly useful in data with spatial, temporal, and spectral dimensions like MRI data and hyperspectral image. CP decomposition has been used to reconstruct images from under-sampled data like brain \cite{li2023learned, hou2020matrix}, cardiac \cite{wu2018multiple} MRI and CT scans \cite{zhang2016tensor}. Using the low-rank nature of medical images it achieves high-quality reconstructions with reduced scan times. In magnetic resonance fingerprinting (MRF) reconstruction, CP decomposition is used as a component in a neural network to learn low-rank tensor priors~\cite{li2023learned}, avoiding computationally expensive SVD on high-dimensional data.  

CP decomposition can help reduce redundancy and has been used for brain MRI denoising~\cite{cao2016tensor, cui2020multidimensional}. By selecting the dominant components in CP decomposition, noise and artifacts in medical images can be effectively removed, preserving only the significant structures or patterns. Variants such as Bayesian CP decomposition have been proposed to recover the de-noised tensor from a noisy tensor by estimating rank-one tensors through a variational Bayesian inference strategy~\cite{cao2016tensor}. CP decomposition has been used in a dental CT super resolution task producing high SNR with robust segmentation quality~\cite{hatvani2018tensor}. It is also used in hyperspectral image unmixing with applications in fluorescence microscopy of retina~\cite{dey2019tensor}. In this work, different excitation wavelengths were assembled as a tensor and a non-negative CP decomposition was performed to obtain low-rank factors. Other applications include cardiac $T_1$ mapping \cite{yaman2019low} and feature extraction for multi-modal data, where a robust coupled CP decomposition was used with an assumption that the tensors shared the first factor marix~\cite{zhao2023robust}. In multiphoton microscopy, CP decomposition has been utilized to extract meaningful components from 3D distribution in disease models such as cancer and fibrosis~\cite{uckermann2020label}.


\paragraph{Tucker decomposition.}
Tucker decomposition expresses a tensor as a core tensor multiplied by factor matrices along each mode. Compared with HOSVD and CP decomposition, Tucker decomposition allows for arbitrary forms of factor matrices, providing a more general representation. It has been widely used for predicting missing information in high-dimensional images, with applications in super-resolution \cite{gui2017brain, jia2022nonconvex, hatvaniy2021single} and reconstruction \cite{wu2018multiple, wang2021spectral, li2018efficient}. The general model can be summarized as $\mathcal{T} = \mathcal{A}\mathcal{X} $ where $T$ is the observed tensor, $X$ is the original image to be predicted and $X$ is a linear operator. By representing $X$ with Tucker decomposition, the operation for each mode and their interactions can be separated, providing interpretable insight into the data. For example, in super-resolution of dental CT images, the high-resolution image is represented by a Tucker decomposition~\cite{hatvaniy2021single}. Applications of Tucker decomposition also includes 
$T_1$ mapping \cite{liu2021accelerating, yaman2019low}. Accelerating the acquisition of 3D $T{1\rho}$ mapping of cartilage, the image tensor is defined as the sum of a sparse component and a low rank tensor. Tucker decomposition is used for tensor low-rank regularization~\cite{liu2021accelerating}. It can also be used for approximating higher quality images and improving $T_1$ mapping performance~\cite{yaman2019low}.  


Another application of Tucker decomposition is in medical image fusion \cite{yin2018tensor, zhang2023tdfusion, chen2024multi}. In these methods, original or processed medical images from diverse modalities are represented with Tucker decomposition sharing the same core tensor or factor matrices. Sparsity regularization is introduced to the core tensors. For example, Yin \textit{et al.}~\cite{yin2018tensor} tackled image fusion by first learning a dictionary from training images. They represented each image as a tensor split into unique sparse core tensors and shared factor matrices. When fusing images, they used these same factor matrices to compute the new tensor representations, letting the core tensors dictate the fusion weights. Beyond MRI, Tucker decomposition has been leveraged to fuse multiphoton imaging data with complementary modalities, such as Raman or fluorescence lifetime imaging, allowing integrative tissue characterization in tumor microenvironments~\cite{karahan2015tensor}. It enables robust feature extraction and pattern recognition, facilitating automated classification of healthy versus diseased tissue regions. 
Besides, Tucker decomposition has shown the potential to be used for medical image segmentation \cite{khan2022deep, weber2025posttraining}, usually as a component of a deep neural network. In knee cartilage MRI segmentation, the original input images are reconstructed by Tucker decomposition with low rank~\cite{khan2022deep}. 
Both the original and reconstructed inputs are fed into the neural network to learn the inter-dimensional tissue structures. In another work~\cite{weber2025posttraining}, the Tucker-decomposed convolution is proposed to replace the conventional 3D convolution and improve computational efficiency of the neural network. 

The application of Tucker decomposition in MRI and multiphoton brain imaging data extends to optogenetics and imaging studies, where simultaneous stimulation and recording of neuronal populations require advanced computational techniqus to disentangle overlapping signals. It separates the evoked responses from spontaneous activity in high-dimensional optogenetic data, aiding the identification of circuit-level changes in neurodevelopmental and neurodegenerative disorders~\cite{erol2022tensors}.

\paragraph{HOSVD.} The applications of HOSVD, a generalization of the matrix SVD, are ubiquitous in biomedical image analysis. As a constraint case of Tucker decomposition, the factor matrices and core tensor of HOSVD are orthogonal. Therefore, HOSVD is less flexible than Tucker decomposition but can ensure decorrelation between components along each mode. For its ability to use the redundancy in high-dimensional data, HOSVD has been used for brain MRI denoising \cite{fu20163d, zhang2017denoise, bustin2019high, wang2020modified, kim2021denoising, olesen2023tensor}. Specifically,
Marchenko-Pastur principal component analysis (MPPCA), a matrix-based method for denoising, is generalized to tensor MPPCA based on HOSVD so that it can directly process multidimensional MRI data~\cite{olesen2023tensor}. A patch-based HOSVD method for denoising has also been proposed, combined with a global HOSVD method to mitigate stripe artifacts in the outputs~\cite{zhang2017denoise}.  
Like Tucker decomposition, HOSVD has been used for medical image reconstruction~\cite{liu2021calibrationless, yi2021joint, roohi2016dynamic, zhang2022compressed} of MRI data by exploiting sparsity and low rank properties. It has also been applied in low-rank tensor approximation of the multi-slice image tensor in parallel MRI reconstruction~\cite{liu2021calibrationless}. HOSVD, analogous to the Tucker decomposition, enables the computation of a core tensor and the factorization of three-way tensors into three matrices. This capability has proven particularly effective for feature extraction, as demonstrated in a blood cell recognition task and wavelet transform in colored pupil images, where they are naturally represented as three-way tensors~\cite{le2023tensor,giap2023adaptive}. 

HOSVD has found utility in analyzing complex spatiotemporal patterns in brain imaging data acquired through single-photon and multiphoton techniques such as two photon and three-photon calcium imaging~\cite{grewe2010high}. It has been applied to denoise neuronal activity maps while preserving underlying spatiotemporal dynamics by reducing motion artificats in calcium imaging datasets. It improved the detection of neuronal ensembles in awake behaving animals~\cite{cho2023robust}.


\paragraph{t-SVD.}
    Unlike HOSVD, which is based on tensor-matrix product, tensor-SVD (t-SVD) represents a 3-way tensor as the \say{t-product} of three 3-way tensors. The t-SVD, a relatively recent and nuanced tensor decomposition technique, has been effectively employed in a range of medical imaging applications, particularly for enforcing low-rank regularization constraints. It has been used for MRI reconstruction \cite{jiang2020improved, liu2025dynamic, ai2018dynamic, liu2023low}. Specifically, it has been used to approximate the rank minimization problem, which is NP-hard to a tensor nuclear norm minimization, with applications in low-rank tensor regularization in a dynamic MRI reconstruction model~\cite{liu2025dynamic}. Similar to other tensor decomposition methods, t-SVD has been used in denoising MRI images~\cite{khaleel2018denoising, kong2017new, khaleel2018denoising2} as well as in image segmentation tasks~\cite{shi2021multi}. A low-rank approximation of the noisy image is obtained by thresholding the t-SVD coefficients in the Fourier domain to perform denoising~\cite{khaleel2018denoising2}. For image segmentation of pathological liver CT, a low-rank tensor decomposition is performed based on t-SVD~\cite{shi2021multi}. Specifically, it is used to recover the underlying low-rank structure of the 3D images and generate tumor-free liver atlases.  
    t-SVD based low-rank approximations have also been used to model neural network connectivity and extract functional components from large-scale neural activity data~\cite{williams2018unsupervised}.

As observed in several tensor decomposition methods applied in biomedical imaging, it is often used together with deep learning techniques such as convolutional neural networks (CNNs)~\cite{yaman2019low,oymak2021learning, khan2022deep, li2023learned}. Deep learning techniques have shown empirical success to process and extract features from images, or to reconstruct them. However, their effectiveness is still a mystery with their heavy dependence on several parameters and their sensitivity to noise along with lack of generalization to out-of-distribution data, and overfitting in limited training samples~\cite{chen2023deep}. Tensor decompositions play a crucial role here extracting meaningful features which increase generalizability of neural network models. Tensor decomposition methods have shown to learn kernels from training data and increase performance of CNNs~\cite{oymak2021learning}. They are also used to reduce training parameters of CNNs and reduce model size, leading to effective training~\cite{liu2023tensor}. Hence, TD methods can be used in conjunction with deep learning methods to enhance the overall throughput and performance of downstream tasks such as prediction, segmentation, reconstruction, super-resolution, etc. in analyzing biomedical images.

\subsection{Analyzing Multi-omics Data}\label{sec:td_multiomics}


The application of tensor decomposition in multi-omics data analysis has significantly advanced in recent years, offering scalable and interpretable solutions for integrating complex biological datasets~\cite{hore2016tensor, amin2023tensor, lee2018gift, taguchi2022adapted, leistico2021epigenomic, taguchi2023tensor, wang2023probabilistic, tsuyuzaki2023sctensor, jung2021monti, taguchi2018tensor, fang2019tightly, taguchi2020tensor}. 
Compared to traditional approaches (e.g., clustering, PCA, correlation analysis), which typically analyze relationships between only one or two variables at a time, TD simultaneously captures complex interactions across multiple omics layers, preserving high-order biological structures~\cite{jung2021monti}. As multi-omics studies continue to generate increasingly high-dimensional and heterogenous data, TD techniques have evolved to address challenges related to dimensionality reduction, data sparsity, and the extraction of biologically meaningful patterns. 

One major development in tensor decomposition for multi-omics datasets has been the integration of probabilistic models to address sparsity and noise. Traditional TD methods assume homogenous data distributions, which limits their ability to model multiomics data with intrinsic variability. Probabilistic TD methods, like \texttt{SCOIT} (Single Cell Multiomics Data Integration with Tensor decomposition)~\cite{10.1093/nar/gkad570}, leverage statistical distributions such as Gaussian, Poisson, and negative binomial models to better capture variability across different omics layers. These models enhance downstream analyses such as cell clustering, gene expression integration, and regulatory network inference, making them particularly valuable for single-cell transcriptomics and epigenomics. To improve biological interpretability, non-negative TD methods enforce non-negativity constraints on the factorized components, ensuring that all contributions to latent factors remain positive. This is particularly useful in biomedical research, where feature contributions must be biologically meaningful. A prime example is \texttt{MONTI} (Multi-Omics Non-negative Tensor Decomposition for Integrative Analysis)~\cite{jung2021monti}, which selects key molecular features by identifying non-negative latent factors that drive disease progression. By preserving additive relationships in multi-omics features, \texttt{MONTI} has been effective in distinguishing molecular signatures associated with cancer subtypes, enhancing patient stratification.

Structural constraints can also be incorporated in TD to improve the separability of multi-omics data while preserving both local and global values. \texttt{BioSTD} (Strong Complementarity Tensor Decomposition Model)~\cite{gao2023biostd} uses t-SVD to factor multi-omics data from the Cancer Genome Atlas (TCGA) and enforces a strong complementarity constraint to maintain high-dimensional spatial relationships across omics layers. This improves coordination between different types of omics data and ensures that redundant features do not obscure biologically significant patterns. Such an approach is particularly beneficial when analyzing high-dimensional datasets in precision oncology, where feature redundancy can compromise the interpretability of predictive biomarkers. Multi-omics integration often faces challenges due to the unequal feature dimensions across omics layers, where transcriptomics datasets typically contain tens of thousands of features, while proteomics and metabolomics datasets are comparatively sparse. \texttt{GSTRPCA} (Irregular Tensor Singular Value Decomposition Model)~\cite{Cui2024GSTRPCA} overcomes this limitation by maintaining the original data structure while incorporating low-rank and sparsity constraints using a tensor-PCA based model. This enables effective feature selection and clustering without requiring aggressive pre-processing or feature imputation, thereby preserving biologically relevant information.

Handling missing values in multi-omics datasets remains a fundamental challenge, particularly in epigenomics, where experimental limitations often result in incomplete data across different cell types. \texttt{PREDICTD} (Parallel Epigenomics Data Imputation with Cloud-Based Tensor Decomposition)
~\cite{Durham2018PREDICTD} addresses this issue by constructing a tensor where dimensions represent cell types, genomic regions, and epigenomic marks, allowing the decomposition process to identify latent factors that capture underlying regulatory patterns. By applying CP decomposition, the method imputes missing values using latent factor reconstruction. This cloud-based framework has been validated in large-scale projects such as ENCODE and the Roadmap Epigenomics Project, highlighting its utility in functional genomics. Although tensor decomposition effectively reduces dimensionality, one of its key challenges is ensuring biological interpretability. Guided and Interpretable Factorization for Tensors~\cite{lee2018gift} (\texttt{GIFT}) enhances interpretability by incorporating functional gene set information as a regularization term within the decomposition process. It adapts Tucker decomposition to factorize with prior biological knowledge, \texttt{GIFT} ensures that extracted components correspond to meaningful gene regulatory patterns. Applied to datasets such as the PanCan12 dataset from TCGA, \texttt{GIFT} has successfully identified relationships between gene expression, DNA methylation, and copy number variations across cancer subtypes, outperforming traditional tensor decomposition approaches in terms of accuracy and scalability. 

In terms of scalability, TD has evolved to accommodate large-scale genomic data through parallelized optimization techniques. An example is \texttt{SNeCT} (Scalable Network Constrained Tucker Decomposition)~\cite{Choi2020SNeCT} which extends traditional Tucker decomposition by incorporating prior biological networks into the decomposition process. This method constructs a tensor from multi-platform genomic data while integrating pathway-based constraints to improve feature selection and classification performance. By leveraging parallel stochastic gradient descent, \texttt{SNeCT} efficiently processes large-scale datasets such as TCGA’s PanCan12 dataset. Its ability to integrate gene association networks ensures that the resulting factor matrices maintain biological relevance, making it particularly effective for cancer subtype classification and patient stratification. Standard TD approaches often treat time as a discrete variable, limiting their ability to model continuous temporal changes in biological processes. Temporal Tensor Decomposition~\cite{Shi2024TEMPTED} (\texttt{TEMPTED}) overcomes this limitation by integrating time as a continuous variable within the decomposition framework. This method enables the characterization of dynamic transcriptional and epigenomic changes across different time points, making it particularly useful for analyzing developmental processes and disease progression. \texttt{TEMPTED} also accounts for non-uniform temporal sampling and missing data, providing a more comprehensive approach to longitudinal multi-omics analysis. More recently, Mitchel \textit{et al.} have introduced a computational approach called single-cell interpretable tensor decomposition (\texttt{scITD}), that utilizes tensor decomposition to analyze single-cell gene expression data~\cite{mitchel2024coordinated}. This approach enables the identification of coordinated transcriptional variations across multiple cell types, facilitating the discovery of common patterns among individuals, in turn facilitating patient stratification.

As tensor decomposition methodologies for multi-omics datasets continue to evolve, future advancements will likely focus on improving real-time data analysis capabilities, refining computational efficiency, and expanding the applicability of TD in emerging fields such as spatial transcriptomics and single-cell multi-omics. The continued refinement of these methods will be critical for advancing precision medicine and uncovering novel insights into complex biological systems.

\subsection{Spatial Transcriptomics}\label{sec:td_spatial}

Spatial transcriptomics enables the mapping of gene expression patterns within tissue architecture, providing critical insights into cellular organization and function. However, analyzing such high-dimensional data remains a computational challenge, especially when integrating multiple regions of interest (ROIs) to uncover spatially organized molecular patterns. As discussed above, tensor decomposition is a powerful approach to extract meaningful latent structures from spatial transcriptomic datasets, facilitating deconvolution of cell–cell interactions, spatial clustering of gene expression, and integration of multimodal spatial omics data. Armingol et al.~\cite{armingol2022context} developed \texttt{Tensor-cell2cell}, a non-negative tensor component analysis method, which is an extension of NMF to higher-order tensors—to analyze cell–cell communication while incorporating spatial context. In their work, a 4D tensor representing cell–cell interactions was reconstructed, and a comprehensive error analysis was subsequently performed. Traditional approaches often rely on bulk transcriptomic deconvolution, which disregards spatial positioning, but \texttt{Tensor-cell2cell} leverages tensor decomposition to preserve high-order interactions between signaling pathways across different tissue regions. By integrating spatially resolved transcriptomic profiles, this method enhances the identification of key cell–cell communication networks that drive tissue function and disease progression. In another study~\cite{broadbent2024deciphering}, a graph-guided Tucker decomposition method was used to decipher high-order structures in spatial transcriptomic data. This approach combines TD with graph-based priors to account for spatial dependencies between neighboring regions of interest. Unlike conventional decomposition techniques, which treat gene expression as independent across space, this method embeds structural tissue information into the decomposition process, leading to improved reconstruction of spatial transcriptomes and more biologically meaningful clustering of tissue subregions. Low-rank TD methods have also been utilized in characterizing the spatiotemporal transcriptome of the human brain~\cite{liu2017characterizing}. By applying tensor factorization, the study extracted low-dimensional representations of gene expression across different brain regions and developmental time points, enabling the identification of major gene expression modules that define functional brain architecture. This method provides a scalable framework for integrating multi-region transcriptomic datasets, offering insights into both spatial organization and temporal dynamics of gene regulation in the brain. Tensor-based spatial transcriptomics analysis was further advanced by the introduction of Graph-Guided Neural Tensor Decomposition (\texttt{GNTD})~\cite{song2023gntd}.  This model integrates spatial and functional relationships by embedding gene co-expression networks into the TD framework. Unlike traditional tensor methods that primarily focus on spatial structure, \texttt{GNTD} combines neural network-based embeddings with graph-guided TD, allowing for the reconstruction of missing transcriptomic signals while preserving biologically relevant spatial interactions. This approach enhances the accuracy of transcriptomic reconstructions in sparsely sampled regions, improving the detection of functional tissue microdomains.

Collectively, these studies highlight the power of TD in integrating and analyzing spatial transcriptomic data across different regions of interest. Hence, tensor-based techniques can uncover latent gene expression patterns, reconstruct missing transcriptomic information, and improve spatially aware clustering of cells. As spatial transcriptomics technologies continue to evolve, tensor decomposition will play a crucial role in enabling more precise and scalable analyses of high-dimensional spatial omics datasets.

\subsection{Challenges in Biomedical Tensor Decomposition}\label{sec:challenges}

\begin{figure}[htbp]
    \centering
    \includegraphics[width=\linewidth]{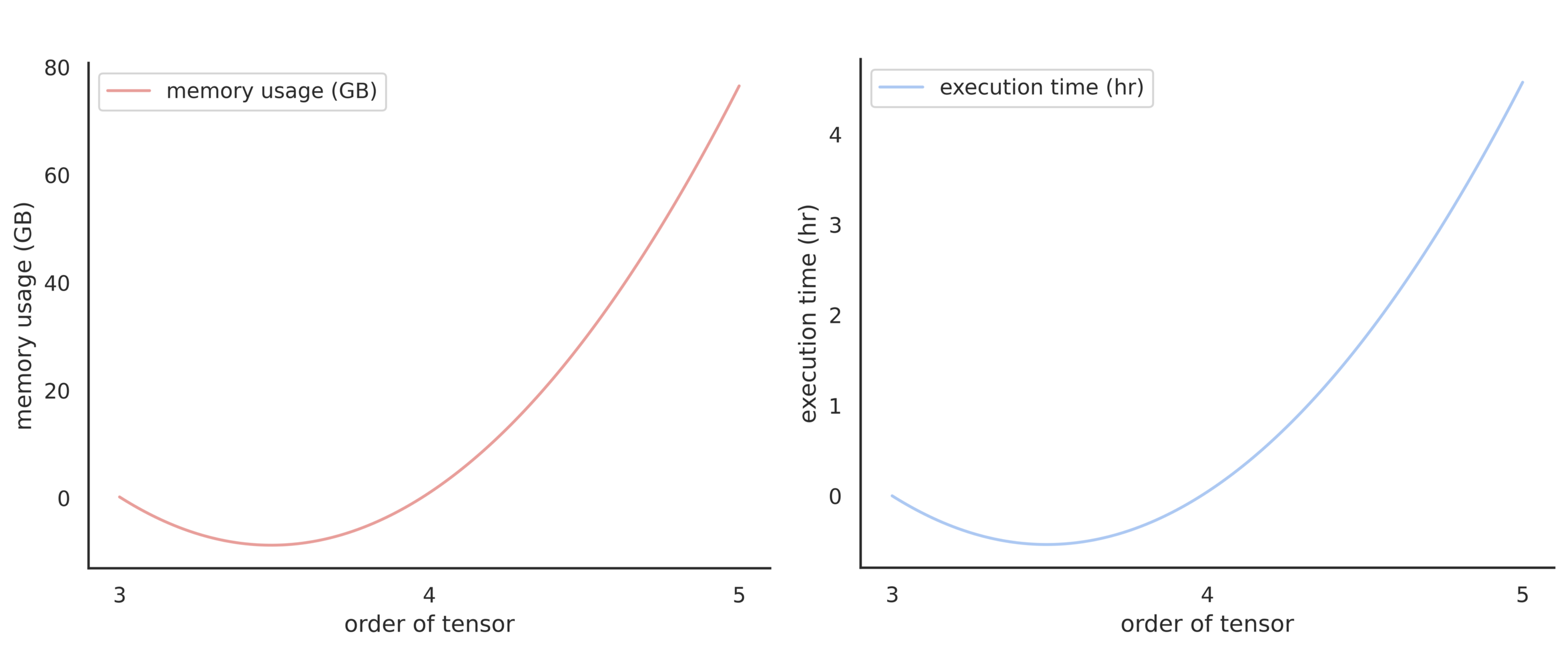}
    \caption{Memory usage (in gigabytes) and wall-clock execution time (in hours) for Tucker decomposition of a random dense tensor of size $N^d$, where $N=100$ and the order $d$ is represented in the x-axis.}
    \label{fig:mem-time}
\end{figure}

Tensor-based modeling for biomedical data presents inherent challenges due to the high-dimensional nature of these datasets, which results in computationally intensive operations. We choose to demonstrate the challenges in light of Tucker decomposition as it is straightforward to comprehend as a combination of a core tensor and factor matrices (Definition~\ref{def:tucker}), and also widely used in biomedical data analysis. There are several factors that governs the time and space complexity of Tucker decomposition, namely $d$, the order of the tensor, $N$, the size of the data, and the size of the core tensor $T$. The average case storage and time complexity scales exponentially with respect to these three parameters, yielding $\mathcal{O}(dNT+T^d)$ and $\mathcal{O}(dN^{d-1}T)$, respectively. We observe that in practice, processing dense tensors with randomly generated data from normal distribution, makes it scale exponentially with increasing order in large-scale high-performance computing clusters (HPC) (Figure~\ref{fig:mem-time}) for both memory and time. We computed the Tucker decomposition in IBM's HPC cluster with Supermicro servers using the \texttt{tensorly}~\cite{kossaifi2019tensorly} package in Python. 
Moreover, there is a critical need to balance the fidelity of the latent representations with computational efficiency, as discussed in Section~\ref{sec:hardness}. In the following sections, we detail these challenges, illustrating the hardness in terms of space and time requirements for processing real-world imaging and multi-omics data as tensors and performing Tucker decomposition. For each of these cases, we also describe how the order of the tensor plays a crucial role in determining tractability of tensor decompositions.

\subsubsection{Biomedical Imaging}
In medical imaging, tensor representations of images seeks to preserve the intrinsic high-dimensional structure of the images. However, computation of numerical results from tensors often requires converting these models into matrix or vector forms. This leads to the construction of large matrices involving Kronecker products, imposing significant computational burdens on current hardware. Consequently, computations are typically divided into small, overlapping patches, an approach that is not universally applicable, especially in medical imaging. For example, in MRI super-resolution, the low-resolution image is derived from a global Fourier transform of the entire high-resolution image, making patch-based processing infeasible. 

\begin{figure}[!htbp]
    \centering
    \includegraphics[width=0.85\linewidth]{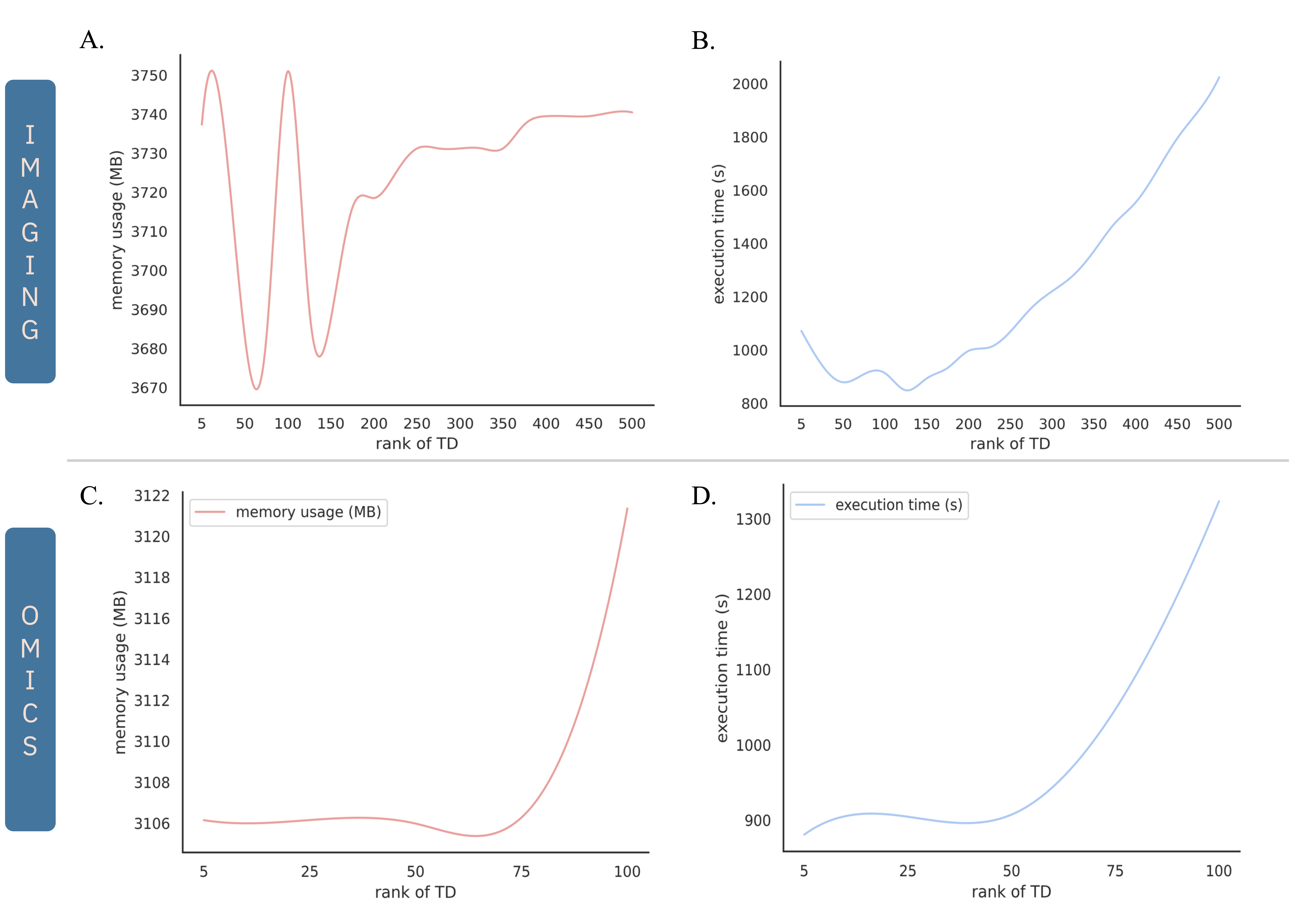}
    \caption{Measure of computational hardness in performing Tucker decomposition on biomedical data: \textbf{A.} Memory usage (in Megabytes) in decomposing 3D MRI data of size ($512 \times 784 \times 912$); \textbf{B.} Wall-clock time (in seconds) of decomposing the 3-way MRI data; \textbf{C.} Memory usage (in Megabytes) in decomposing a 3-way multi-omics tensor of size ($281 \times 10000 \times 100)$; \textbf{B.} Wall-clock time (in seconds) of decomposing the 3-way multi-omics tensor.}
    \label{fig:hardness}
\end{figure}

Another general challenge in tensor decomposition is rank selection. For example, in Tucker decomposition, selecting the size of the core tensor, which is also the rank, is hard. It is significant since it determines the level of dimensionality reduction and the trade-off between performance and computational efficiency. If the rank is too low, capturing the latent structures in the data may be weakened, while a high rank increases computational complexity and memory usage significantly. As we discuss in Section~\ref{sec:hardness}, understanding the optimal rank of the tensor leads to a low MMSE, yielding efficient signal recovery. This phenomenon is true for all tensor decomposition techniques and their applications in medical imaging ranging from MRI to neural activity patterns and calcium imaging. Using multi-slice MRI data with image size $512\times784\times912$ and voxel size $0.2\times0.18\times0.18$~mm, where the dimensions represent the number of slices (512), and the in-plane resolution ($784\times912$) of the imaging sequence, we performed Tucker decomposition to obtain the latent factors representing the image. We observed an exponential trend in the time to compute ranks up to 500 mimicking the performance of the dense third-order tensor earlier in Figure~\ref{fig:mem-time}. However, an oscillating memory usage was recorded when increasing the rank of the tensor decomposition, possibly due to discrete memory allocations, triggering change in memory block sizes and internal data structures (Figure~\ref{fig:hardness} A and B). 

Besides, real-world tensors in medical imaging, often contain noise or artifacts that reduce performance and increase computational requirements for robust solutions. Performance reduction when adding noise has been reported in different methods using CP decomposition \cite{zhao2023robust}, Tucker decomposition \cite{prevost2020hyperspectral}, t-SVD \cite{liu2025dynamic}.

Applications of TD methods in tissue and brain imaging also leads to significant challenges, particularly, in the context of high-speed multiphoton microscopy and large-scale neural recordings. These challenges stem from the high-dimensional and multi-modal nature of the data acquired from multiphoton imaging techniques, such as second and third harmonic generation microscopy, two-photon and three-photon imaging, and functional calcium imaging. These datasets are inherently large, spanning spatial, temporal, and spectral domains, making tensor-based approaches computationally demanding~\cite{kolda2009tensor}. Another critical challenge is motion artifacts and signal contamination in live brain imaging. In awake behaving animals, head movements, heartbeat, and breathing introduce artifacts in calcium imaging and optogenetic recordings, complicating the extraction of meaningful neural signals. While tensor-based motion correction techniques have shown promise, they often require large computational resources and can introduce biases if the decomposition fails to properly separate artifacts from genuine neural activity~\cite{kara2024facilitating}. 
Furthermore, biological noise and tissue heterogeneity present difficulties in tensor-based models applied to tissue imaging. In label-free multiphoton microscopy, where contrast arises from intrinsic tissue properties rather than fluorescent markers, non-uniform signal intensities, variations in optical scattering, and background noise can degrade the performance of decomposition techniques~\cite{borile2021label}. Robust pre-processing and artifact rejection methods are essential to improve the reliability of tensor-based segmentation, classification, and feature extraction in histopathological and live-tissue imaging. Finally, hardware limitations pose another barrier. While parallel computing and GPU acceleration have improved computational feasibility, real-time applications, such as closed-loop optogenetics or high-speed volumetric imaging, require further advancements in tensor decomposition algorithms to achieve near-instantaneous data processing and decision-making. Addressing these challenges will be crucial for fully harnessing tensor decomposition in large-scale tissue and brain imaging studies.

\subsubsection{Multi-omics Analysis}
Multi-omics data analysis presents several challenges, particularly in handling, integration, and interpretation. One of the major difficulties lies in the computational complexities associated with large-scale multi-omics datasets. These datasets are often high-dimensional due to the large number of features across multi-omics layers. Finding an optimal rank in this case is equally challenging as with imaging data. 
To illustrate these computational challenges, we performed Tucker decomposition on a multi-omics dataset combining autosomal single nucleotide polymorphisms and metabolites from germline PTEN Hamartoma Tumor Syndrome carriers with a tensor size of (264, 10000, 100), where the dimensions represent the number of samples (264), genomic features (10,000), and metabolomic features (100). Figure~\ref{fig:hardness} C shows memory requirements (in Megabytes) as the decomposition rank varies from 5 to 100, revealing a constant trend before exponentially increasing after rank 75. The second plot (Figure~\ref{fig:hardness}D) illustrates wall-clock time (in seconds), which steadily rises with increasing rank of the Tucker decomposition. These results highlight a critical trade-off between rank selection and computational efficiency, emphasizing the need for optimized feature selection strategies to balance computational feasibility and biological relevance when applying tensor decomposition to multi-omics data. 

Another key challenge is the heterogeneity of multi-omics data. These datasets comprise diverse data types, including continuous, binary, and categorical variables, complicating integration efforts. This heterogeneity complicates the integration process, as traditional tensor decomposition methods such as Tucker and CANDECOMP / PARAFAC (CP) may not adequately model complex interactions and different data types~\cite{xu2015bayesian}. Beyond computational and integration challenges, interpreting the latent factors extracted from tensor decomposition in a biologically relevant context can be challenging. Moreover, multi-omics data are often noisy and contain outliers. This can distort the results of the tensor decomposition. However, there are new methods such as \texttt{SCOIT}~\cite{10.1093/nar/gkad570} which attempt to address this challenge by incorporating various distributions to model noise and sparsity in the data. Finally, multi-omics data often contain missing values or sparse entries, which can significantly impact the performance of tensor decomposition methods. Several traditional approaches require complete data or extensive preprocessing to handle missing data, which can be time-consuming and may introduce biases \cite{xu2015bayesian, taguchi2021tensor}. While tensor decomposition provides a powerful framework for multi-omics integration,  overcoming computational constraints, handling data heterogeneity, improving interpretability, mitigating noise, and addressing missing data challenges remain critical areas for future research and methodological advancements.


\subsubsection{Spatial Transcriptomics}
Despite the advantages of tensor decomposition in spatial transcriptomics, several challenges remain in its application to high-dimensional and spatially heterogeneous datasets. One key challenge is the complexity of spatial dependencies, as gene expression varies not only across different tissue regions but also within microenvironments, making it difficult to define an optimal tensor structure that captures both local and global expression patterns. Traditional tensor decomposition methods often assume a predefined rank or spatial resolution, which may not be suitable for highly dynamic and heterogeneous tissue architectures. Additionally, sparsity in spatial transcriptomic data—where many genes are not detected in all spatial locations—limits the accuracy of decomposition models and can lead to biased reconstructions, especially when missing data is not randomly distributed but influenced by biological or technical factors. Another challenge is computational scalability, as large-scale spatial transcriptomics datasets with thousands of genes and spatial positions require substantial memory and processing power, making traditional tensor-based approaches computationally expensive. Furthermore, biological interpretability of tensor decomposition outputs remains a challenge, as extracted latent components may not always correspond to clear biological pathways or cell-type-specific interactions, necessitating additional validation with external datasets or functional assays. Finally, integration of multimodal spatial omics data poses a challenge, as combining spatial transcriptomics with proteomics or epigenomic data using tensor methods requires designing decomposition strategies that can accommodate different data modalities, resolutions, and measurement biases. Addressing these challenges will be crucial for unlocking the full potential of tensor decomposition in spatial transcriptomics and improving its utility for biological discovery.

\section{Quantum Algorithm for Tensor Decomposition}~\label{sec:td_quantum} 
Quantum computing leverages principles like superposition, entanglement, measurement, and destructive interference among solution probabilities to perform computational tasks more efficiently than classical computing (See Section 2 of \cite{basu2023towards} for a short primer on quantum computing and \cite{nielsen2010quantum} for a detailed introduction). In the past decade, quantum algorithms for matrix and tensor decomposition have emerged as powerful tools with significant applications in machine learning, data compression, quantum chemistry, and more. A notable advancement is the quantum singular value decomposition (QSVD), which expresses a matrix $A$ as $A = U \Sigma V^\dagger$, where $U$ and $V$ are unitary matrices and $\Sigma$ is a diagonal matrix of singular values. A quantum algorithm for the singular value decomposition of nonsparse low-rank matrices, providing exponential speedup over classical algorithms under certain conditions was proposed by Rebenstrot \textit{et al.}~\cite{rebentrost2018quantum}. This algorithm employs quantum state preparation techniques and Quantum Phase Estimation (QPE) to efficiently extract singular values and corresponding singular vectors.

Building on this, variational quantum singular value decomposition (VQSVD) algorithm was introduced~\cite{wang2021variational}. Designed for PFTQD, VQSVD uses a hybrid quantum-classical optimization loop to approximate singular values and vectors by minimizing a cost function. This makes it practical for current quantum hardware and effective for matrices where only a few singular values are significant.
Another significant development is quantum principal component analysis (QPCA)~\cite{lloyd2014quantum} introduced a quantum algorithm for PCA that can extract principal components exponentially faster than classical algorithms under certain conditions . The QPCA algorithm uses quantum algorithms for Hamiltonian simulation and phase estimation to find the eigenvalues and eigenvectors of the covariance matrix, processing large datasets encoded in quantum states and making it promising for big data analysis. Foundational to many quantum linear algebra algorithms is the quantum algorithm for solving linear systems of equations developed by Harrow, Hassidim, and Lloyd~\cite{harrow2009quantum}, also known as the HHL algorithm. The HHL algorithm solves systems of the form $Ax = b$ in logarithmic time relative to the size of the system under conditions such as sparsity and a low condition number of matrix $A$. It is crucial for quantum matrix computations, including inversions and decompositions, with applications spanning machine learning and optimization.

Extending these concepts to tensors, which generalize matrices to higher dimensions, offers significant potential in processing multidimensional data. Hastings~\cite{hastings2020classical} introduced classical and quantum algorithms for tensor principal component analysis (Tensor PCA), providing insights into the computational complexity of the problem and demonstrating potential quantum advantages for the problem of spiked tensor decomposition (Equation~\ref{eq:st}). The quantum algorithm presented in \cite{hastings2020classical} proves a quartic speedup over the best classical spectral algorithm. This quantum algorithm takes advantage of QPE and amplitude amplification, coupled with a suitable choice of the input state initialization to achieve the speedup. The quantum Hamiltonian constructed from the initial tensor $T_0$ operates on a set of $n_{bos}$ qudits of dimension $N$. More precisely, for a given tensor $T$ of order $p$ and dimension $N$, the Hamiltonian is a linear operator on the vector space $(\mathbb{R}^N)^{\otimes n_{bos}}$ or $(\mathbb{C}^N)^{\otimes n_{bos}}$ where $n_{bos} \geq \frac{p}{2}$. The bosonic Hamiltonian on the full Hilbert space has basis elements of the form $|\mu_1\rangle \otimes |\mu_2\rangle \otimes \dots \otimes |\mu_{n_{bos}} \rangle$ where each $\mu_i \in \{0, 1, \dots , N-1\}$ and can be written as,

\begin{equation}\label{eq:hamiltonian}
H(T) = \frac{1}{2} \sum_{i_1, \dots , i_{p/2}} \left( \sum_{\mu_1, \dots , \mu_p} T_{\mu_1, \mu_2, \dots, \mu_p} |\mu_1\rangle _{i_1} \langle \mu_{1+p/2} |\otimes |\mu_2\rangle _{i_2} \langle \mu_{2+p/2} |\otimes \cdots \otimes |\mu_{p/2}\rangle _{i_{p/2}} \langle \mu_{p}| + h.c.  \right)
\end{equation}

where the first sum iterates over unique set of qudits $i_1, i_2, \dots, i_p$, $T_{\mu_1, \mu_2, \dots, \mu_p}$ are the corresponding elements from the tensor $T$ and $h.c.$ refers to adding Hermitian conjugates for the terms. Outer products $|\mu_n\rangle _{k} \langle\mu_{m}|$ are performed on the corresponding qudit $k$. Then the quantum algorithm in \cite{hastings2020classical} recovers the leading eigenvalue and the corresponding eigenvector, assuming the detection is successful, and outputs a vector that has large normalized overlap with the signal vector $v_{sig}$. 

The Hamiltonian $H(T_0)$, inherits its spectral properties, which in turn dictate the sample complexity of the quantum algorithm. A key factor is the spectral gap $\Delta$, which determines how well the ground state can be distinguished from excited states. Since the largest eigenvalue of $H(T_0)$ is driven by the signal term $\lambda v_{\text{sig}}^{\otimes p}$, a higher SNR (larger $\lambda$) results in a larger spectral gap, making it easier to resolve the ground state with quantum phase estimation and reducing sample complexity, which scales as $\mathcal{O}(1/\Delta^2)$. However, when the SNR is low, the signal eigenvalue becomes buried in the noise spectrum, causing the spectral gap to shrink and significantly increasing the number of measurements required. Additionally, the operator norm $\|\hat{H}\|$, which grows with the tensor order $p$ and dimension $N$, influences the spread of energy levels and dictates the precision needed in phase estimation, affecting the total measurement overhead. While the quantum algorithm achieves a quartic speedup over classical methods, its efficiency is still constrained by the interplay between SNR, tensor order, and spectral properties - if the spectral gap is too small due to low SNR or high $p$, the measurement complexity increases, limiting the quantum advantage.

In QPE, ancilla qubits store the binary representation of the estimated eigenvalue, determining the precision of the approximation. It also determines the depth of the quantum circuit. Quantum Amplitude Amplification (the second step in Hastings' algorithm~\cite{hastings2020classical}) improves the success probability of measuring the correct eigenvalue but does not alter the role or requirement of ancilla qubits in QPE. The number of ancilla qubits, as well as the number of times the controlled unitary is applied to the main register, is dictated by the precision requirement of the problem. Specifically, to achieve 2, 4, 6, and 8 decimal digit precision, the required number of ancilla qubits is 7, 14, 20, and 27, respectively. It also means that the controlled unitary on the main register will be applied $\sim 2$, $\sim2^3$, $\sim 2^5$, and $\sim 2^7$  times respectively. Such depth makes an underlying quantum error correcting layer necessary for the algorithm. In practice, when applied on biomedical data, up to four decimal digit precision is often sufficient to qualitatively evaluate tasks such as clustering, prediction, classification, or feature extraction. Hence, in realistic terms, a QPE when applied on real-world data would approximately require upto 14 ancilla qubits. However, this estimate is dependent on other factors such as number of qubits, size and order of the tensor, etc.

The quantum algorithmic resources needed to implement this algorithm is high due to QPE yielding high circuit depths, however a more recent work \cite{zhou2024statistical} provides numerical estimations and analyzes the asymptotic behavior of this model using a $p$-step quantum approximate optimization algorithm (QAOA), making it more amenable in near-term utility-scale quantum computers.

While explicit quantum algorithms for CP (CANDECOMP/PARAFAC) and Tucker decompositions are still under development, the principles of quantum linear algebra can be extended to these tensor factorizations. This extension potentially offers speedups in processing high-dimensional data, which is crucial in fields like computer vision, chemometrics, neuroscience, and quantum chemistry. As quantum computing technology advances, these algorithms pave the way for more efficient data processing techniques and inspire further research into quantum algorithms for complex data structures.

\section{Quantum Tensor Decomposition for Biomedical Data}~\label{sec:td_usecase} 
\begin{figure}
    \centering
    \includegraphics[width=\linewidth]{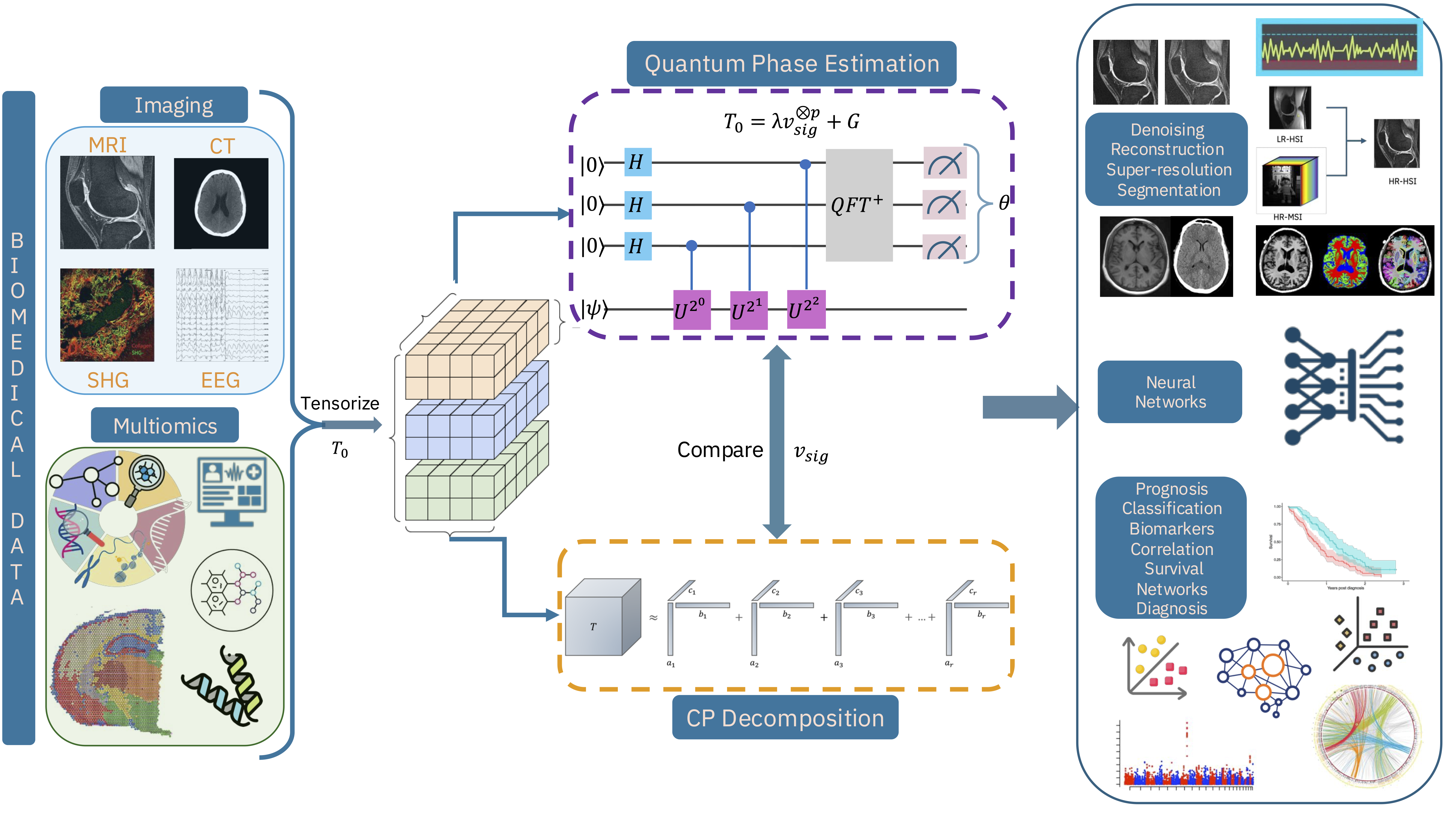}
    \caption{Proposed framework of quantum tensor decomposition methods for analyzing biomedical data for various downstream tasks.}
    \label{fig:qtd_framework}
\end{figure}
The applications of quantum tensor decomposition for analyzing biomedical data must integrate the recent advances in QTD algorithms~\cite{hastings2020classical,zhou2024statistical} with existing approaches in classical algorithms in spiked tensor decomposition or CP decomposition. The QTD algorithm should be compared against these classical approaches for quantitative and qualitative benchmarking and assessing the problems that are more suitable for the quantum algorithm. We propose a framework to implement QTD in biomedical data analysis (Figure~\ref{fig:qtd_framework}) and perform a host of downstream tasks targeted towards the nature and modality of the data.  

The framework ingests multi-modal data such as imaging or multi-omics including spatial transcriptomics. It tensorizes them into higher-order tensors (with orders greater than two) to produce the tensor $T_0$ in the form of a spike tensor problem (Equation~\ref{eq:st}). Then we take a multi-pronged approach to find the underlying signal $v_{sig}$. We form the Hamiltonian, $H(T_0)$ from $T_0$ as per Equation~\ref{eq:hamiltonian} and perform a spectral decomposition, as shown in previous work~\cite{hastings2020classical}, which proposed both classical and quantum algorithms for spectral decomposition. We decompose $H(T_0)$ by solving a QPE problem~\cite{kitaev1995quantum} in FTQD, to efficiently extract eigenvalues and corresponding eigenvectors. The QPE circuit is shown in the inset of Figure~\ref{fig:qtd_framework}'s quantum approach. Once we obtain the leading eigenvalue with its corresponding eigenvector, we can recover $\bar{v}_{sig}$ up to some accuracy. Simultaneously, we can also apply a CP decomposition to the original $T_0$ and obtain the $v_{sig}$. Then, we need to compare the correlation between $\bar{v}_{sig}$ and the $v_{sig}$, to assess the quality of the signal obtained from the quantum algorithm. 

\begin{figure}[H]
    \centering
    \includegraphics[width=0.9\textwidth]{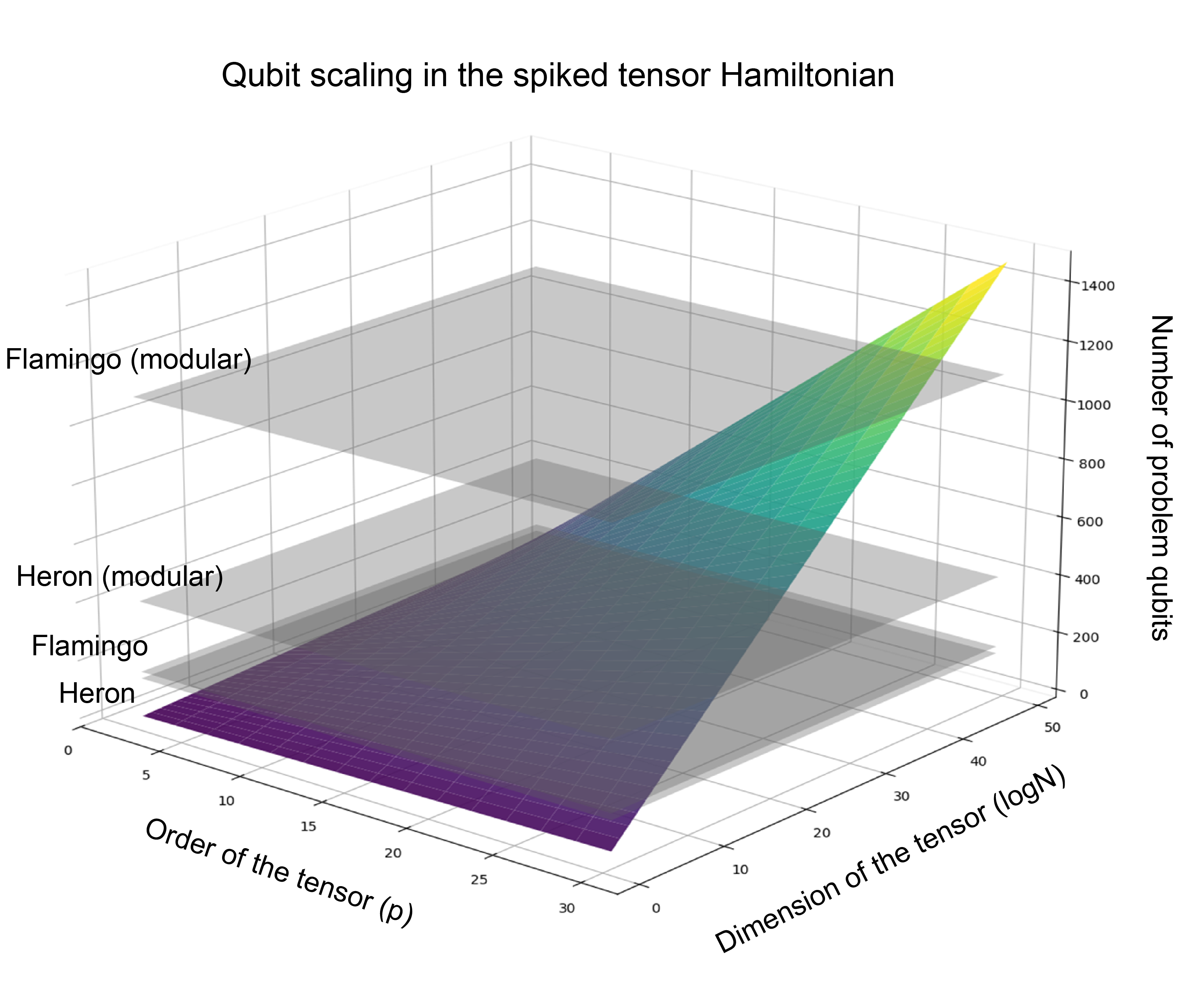}
    \caption{Figure shows the qubit scaling in the spiked tensor Hamiltonian as the tensor order $p$ and dimension $N$ (in log scale) increases. The horizontal planes represent the cut-off points for quantum computers, from IBM Quantum's roadmap, with corresponding number of qubits. In particular, Heron and Flamingo QPUs contain 133 and 156 qubits. With the modular structures where multiple QPUs are coupled, the qubit numbers increase to 399 and 1092 respectively. It is important to note that, our qubit counts are problem qubit counts while the qubit numbers on IBM Quantum roadmap are physical qubits. Depending on the type of algorithm, they may not map to each other on a 1-to-1 basis. Further information can be found at \href{https://www.ibm.com/quantum/technology}{IBM Quantum Technology Roadmap}~\cite{gambetta2020ibm}.}
    \label{fig:qubits}
\end{figure}

Furthermore, to understand the complexity of the problem and how it maps to a quantum computer, we also analyze the number of qubits required to map a given tensor to a qubit Hamiltonian (see Figure \ref{fig:qubits}). Note that the Hamiltonian from Equation \ref{eq:hamiltonian} is formulated for a bosonic system, i.e. an $N$-level qudit system. In Figure \ref{fig:qubits}, we show the final scaling of how this qudit system maps to qubits, since the quantum computers we consider operate natively with qubits. We are aware that there many other parameters to consider in this case, such as the circuit depth (in particular 2-qubit gate depth), overhead from error mitigation/suppression in utility-scale quantum computer or the overhead of error correction for QPE, the trade-off between accuracy and number of repetitions in QPE, the strength of the signal-to-noise ratio in the spiked tensor and the number of bosonic modes required to successfully recover the signal vector etc. However, these require a deeper analysis and are beyond the scope of this paper.  

With this recovered signal, we can perform several downstream tasks in biomedical data analysis. For example, QTD can perform tasks such as denoising, reconstruction, and segmentation of biomedical images. On the other hand, QTD can produce latent representations by integrating multi-modal data in lower dimensions which can be used as input to deep neural networks for downstream analysis. Similarly, these lower-dimensional embeddings can be directly used to discover biomarkers, prognosis, and perform classification or regression tasks with multi-omics data.

\section{Conclusion}~\label{sec:conclusion}
Tensor decomposition methods are highly effective in analyzing multi-modal and multidimensional data such as biomedical data~\cite{cichocki2016tensor}. In this paper we reviewed the state-of-the-art algorithms in TD and analyzed the best practices for implementing them in solving various tasks in biomedical imaging, multi-omics data, and spatial transcriptomics. We designed an unsupervised approach using topic modeling leveraging the transformer architecture with BERTopic~\cite{grootendorst2022bertopic} and identified 536 documents from PubMed in the last decade with applications of TD in biomedicine. We obtained several relevant topics such as on brain EEG, MRI denoising, single-cell omics, transcriptomics, etc. after analyzing and visualizing the literature from PubMed. To further investigate on each of these topics, we performed a systematic review on each sub-domain of biomedical imaging, multi-omics analysis, and spatial transcriptomics, and revealed how TD methods are applied and which tasks are more likely to be solved by tensorizing and obtaining latent factors. For each sub-domain, we also investigated the challenges and limitations of TD, along with a thorough understanding of computational complexity and hardness of the problem. Specifically, we sought to understand the time and space complexity with respect to the growing size of the data to make TD amenable for biobank-scale data. 

Armed with deeper understanding of challenges in applications of TD in biomedicine, we proposed an adaptation of recent advances in quantum computing to develop algorithms for tensor decomposition~\cite{hastings2020classical,zhou2024statistical} for current quantum hardware implementation, with a future outlook on fault-tolerant quantum devices. We proposed a framework (Figure~\ref{fig:qtd_framework}) to apply quantum tensor decomposition and analyze biomedical data to perform various downstream tasks. We acknowledge that there are various challenges in implementing quantum algorithms in PFTQD, due to the parametric nature of the variational quantum algorithms, which need extensive hyperparameter tuning. Furthermore, current quantum computers suffer from hardware noise that can impact computations, leading to instabilities. However, with recent advances in noise mitigation and suppression techniques, we aim to obtain high-quality and robust results from quantum computers in performing the QTD task, at scale. We performed a preliminary resource estimation analysis of implementing QTD in PFTQD using QPE algorithms (Figure~\ref{fig:qubits}) and provide a reaistic roadmap of applying QTD in biomedical data. Integrating quantum algorithms for tensor decompositions with their classical counterparts not only aims to enhance the efficiency of performing TD at scale, but also improve the quality and generalizability of the latent tensor factors in lower dimensions. Quantum computing is an emerging technology which is advancing rapidly with theoretical and practical developments. This comprehensive review of tensor decomposition methods, combined with the proposed development of a quantum tensor decomposition framework, may serve as a model for how quantum computing can influence and drive impact in biomedicine. 

\section*{Acknowledgement}
The authors like to thank Brian Quanz, Vaibhaw Kumar, Charles Chung, Gopal Karemore, and Travis Scholten from IBM Quantum for their insightful feedback.

\newpage 
\bibliographystyle{unsrt}
\bibliography{references}

\newpage
\appendix

\section{ PARAFAC2 as a special case of Tucker decomposition}
    Following example shows that PARAFAC2 decomposition is a specialization of Tucker decomposition. Consider a 3-order tensor \( \mathcal{X} \in \mathbb{R}^{I \times J_k \times K} \), where the size of the second mode (\( J_k \)) varies across slices \( k \). For example:
    \[
    \mathcal{X}(:,:,1) \in \mathbb{R}^{3 \times 2 \times 1}, \quad
    \mathcal{X}(:,:,2) \in \mathbb{R}^{3 \times 3 \times 1}.
    \]
    Here:
    \begin{itemize}
        \item \( I = 3 \) (fixed first dimension),
        \item \( J_k \) varies (\( J_1 = 2, J_2 = 3 \)),
        \item \( K = 2 \) (fixed third dimension).
    \end{itemize}
    
    This varying second mode makes \textbf{PARAFAC2 applicable}, while Tucker decomposition remains a general alternative.
    
    PARAFAC2 models this tensor as:
    \[
    \mathcal{X}_k = \mathbf{A} \, \text{diag}(\mathbf{c}_k) \, \mathbf{B}_k^\top,
    \]
    where:
    \begin{itemize}
        \item \( \mathbf{A} \in \mathbb{R}^{I \times R} \) (shared across slices),
        \item \( \mathbf{c}_k \in \mathbb{R}^{R} \) (slice-specific scaling),
        \item \( \mathbf{B}_k \in \mathbb{R}^{J_k \times R} \) (slice-specific factor matrix),
    \end{itemize}
    and \( \mathbf{B}_k^\top \mathbf{B}_k = \mathbf{B}_j^\top \mathbf{B}_j \) for all \( k, j \). 
    



\section{DEDICOM as a Special Case of PARAFAC}
\label{sec:dedicom-to-parafac}

The DEDICOM model can be seen as a constrained version of the PARAFAC decomposition where:
\begin{enumerate}
    \item The same factor matrix \( A \) is used in both the first and second modes.
    \item An asymmetric relationship matrix \( R \) governs interactions between components.
    \item Diagonal matrices \( D_k \) provide mode-3 scaling.
\end{enumerate}

PARAFAC Decomposition (Section 3 of \cite{kolda2009tensor}):

The PARAFAC (or CP) decomposition for a third-order tensor \( \mathcal{X} \in \mathbb{R}^{I \times J \times K} \) is expressed as:
\begin{equation}
    \mathcal{X} \approx \sum_{r=1}^{R} a_r \circ b_r \circ c_r,
\end{equation}
where \( a_r \in \mathbb{R}^I \), \( b_r \in \mathbb{R}^J \), and \( c_r \in \mathbb{R}^K \) are factor vectors for each mode, and \( \circ \) denotes the outer product.

In elementwise form:
\begin{equation}
    x_{ijk} \approx \sum_{r=1}^{R} a_{ir} b_{jr} c_{kr}.
\end{equation}

DEDICOM Decomposition (Section 5.4 of \cite{kolda2009tensor}):

The three-way DEDICOM model for a tensor \( \mathcal{X} \in \mathbb{R}^{I \times I \times K} \), assuming square slices in mode-1 and mode-2, is given by:
\begin{equation}
    \mathcal{X}_k \approx A D_k R D_k A^\top, \quad \text{for } k = 1, \dots, K,
\end{equation}
where:
\begin{itemize}
    \item \( A \in \mathbb{R}^{I \times R} \) is the factor matrix describing latent components.
    \item \( R \in \mathbb{R}^{R \times R} \) captures asymmetric relationships between components.
    \item \( D_k \in \mathbb{R}^{R \times R} \) is a diagonal matrix representing component weights for slice \( k \).
\end{itemize}

Rewriting the DEDICOM model in elementwise form:
\begin{equation}
    x_{ijk} = \sum_{p=1}^{R} \sum_{q=1}^{R} a_{ip} (D_k)_{pp} r_{pq} (D_k)_{qq} a_{jq}.
\end{equation}

Define \( c_{kr} = (D_k)_{rr} \), the diagonal entries of \( D_k \). Then:
\begin{equation}
    x_{ijk} = \sum_{p=1}^{R} \sum_{q=1}^{R} a_{ip} r_{pq} a_{jq} c_{kp} c_{kq}.
\end{equation}

This structure resembles the PARAFAC model but with:
\begin{itemize}
    \item Shared Factors: The first two modes share the same factor matrix \( A \).
    \item Asymmetry: The relationship matrix \( R \) introduces asymmetry, which is not typically present in PARAFAC.
    \item Scaling: The diagonal matrices \( D_k \) impose additional structure through \( c_{kr} \).
\end{itemize}

\section{INDSCAL as a special case of CANDELINC}
\label{sec:indscal-to-candelinc}
INDSCAL Model (Section 5.1 of \cite{kolda2009tensor}):

   \begin{equation}
       \mathcal{X} \approx \langle A, A, C \rangle = \sum_{r=1}^{R} a_r \circ a_r \circ c_r,
   \end{equation}
   where:

   \begin{enumerate}
       \item $A \in \mathbb{R}^{I \times R}$ is the factor matrix shared across the first two modes (symmetry is enforced).
       \item $C \in \mathbb{R}^{K \times R}$ represents the third mode's variation.
   \end{enumerate}

The symmetry implies $x_{ijk} = x_{jik}$ for all $i$, $j$, and $k$.

CANDELINC Model (Section 5.3 of \cite{kolda2009tensor}):
   \begin{equation}
       \mathcal{X} \approx \langle \Phi_A A, \Phi_B B, \Phi_C C \rangle,
   \end{equation}
   where:

   \begin{enumerate}
       \item $\Phi_A, \Phi_B, \Phi_C$ are constraint matrices.
       \item For example, $\Phi_A A$ represents that the factor matrix $A$ is constrained to lie in the subspace defined by $\Phi_A$.
   \end{enumerate}

The mapping of INDSCAL onto CANDELINC is given below.
\begin{enumerate}
    \item  In the INDSCAL model, the symmetry between the first two modes enforces that $B = A$.
    \item This can be written as a **linear constraint** in the CANDELINC framework: $\Phi_B = \Phi_A = I, \quad B = A$.
  \item Substituting these constraints into the CANDELINC model gives:
  \begin{equation}
      \mathcal{X} \approx \langle \Phi_A A, \Phi_A A, \Phi_C C \rangle.
  \end{equation}
\end{enumerate}

In CANDELINC, the linear constraints are applied via matrices $\Phi_A, \Phi_B, \Phi_C$:

\begin{enumerate}
    \item Setting $\Phi_B = \Phi_A = I$ ensures that $A$ is shared across the first two modes.
    \item The matrix $\Phi_C$ allows $C$ to remain unconstrained, capturing the individual differences specific to the third mode.
\end{enumerate}

CANDELINC accommodates a broader range of linear constraints through $\Phi_A, \Phi_B, \Phi_C$. By specifying: $\Phi_B = \Phi_A = I \quad \text{and enforcing symmetry on } A$,
the INDSCAL model is obtained. Thus, INDSCAL is a specialized case of CANDELINC with symmetry and shared structure constraints.

\section{Topic Modeling Details}

\setcounter{table}{1}
\begin{table}[h!]
\centering
 \begin{tabular}{||c | c||} 
 \hline
 \textbf{Search Term} & \textbf{No. of Documents} \\ 
 \hline\hline
genomics & 46 \\ 
 \hline
transcriptomics & 31 \\ 
 \hline
proteomics & 4 \\ 
 \hline
metabolomics & 3 \\ 
 \hline
epigenomics & 6 \\ 
 \hline
microbiomics & 3 \\ 
 \hline
multiomics & 27 \\ 
 \hline
cancer & 81 \\ 
 \hline
cardiovascular disease & 21 \\ 
 \hline
diabetes & 8 \\ 
 \hline
alzheimer's disease & 20 \\ 
 \hline
neurological disorder & 55 \\ 
 \hline
autoimmune disease & 2 \\ 
 \hline
kidney disease & 4 \\ 
 \hline
obesity & 4 \\ 
 \hline
medical imaging & 218 \\ 
 \hline
 \end{tabular}
 \caption{Number of documents extracted from PubMed for each search term. Each term was used along with ``\texttt{AND} tensor decomposition".}
\end{table}

\begin{table}[h!]
\centering
 \begin{tabular}{||c | c||} 
 \hline
 \textbf{Topic Representation} & \textbf{No. of Documents} \\ 
 \hline\hline
cell\_omics\_expression\_single & 144 \\
\hline
diffusion\_images\_imaging\_noise & 129 \\
\hline
brain\_connectivity\_functional\_eeg	& 108 \\
\hline
segmentation\_dti\_imaging\_95 & 42 \\
\hline
imaging\_bayesian\_high\_neuroimaging & 17 \\
\hline
ecg\_avf\_cardiac\_flow	& 16 \\
\hline
tissue\_mrsi\_matrix\_ncpd & 14 \\
\hline
pd\_ad\_task\_progression & 13 \\
\hline
kidney\_expression\_disease\_cell & 11 \\
 \hline
 \end{tabular}
 \caption{Number of documents per topic embedding.}
\end{table}

\end{document}